\documentclass[pra,aps,twocolumn,eqsecnum,showpacs]{revtex4}

\usepackage{amsmath}
\usepackage{graphicx}
\usepackage{hyperref}
\usepackage{amssymb,amsmath}
\usepackage{epstopdf}
\usepackage{color}
\usepackage{ulem}

\begin{document}

\title{Electromagnetically induced transparency in inhomogeneously broadened \\$\Lambda$-transition with multiple excited levels}

\author{O.S.~Mishina$^{1}$\email{oxana.mishina@spectro.jussieu.fr}, M.~Scherman$^1$, P.~Lombardi$^1$, J.~Ortalo$^1$, D.~Felinto$^2$, A.S.~Sheremet$^3$, A.~Bramati$^{1}$, D.V.~Kupriyanov$^3$, J.~Laurat$^1$, and E.~Giacobino$^1$}
\affiliation{
$^{1}$Laboratoire Kastler Brossel,
Universit\'{e} Pierre et Marie Curie, Ecole Normale Sup\'{e}rieure,
CNRS, Case 74, 4 place Jussieu, 75252 Paris Cedex 05, France\\
$^{2}$Departamento de Fisica, Universidade Federal de Pernambuco, 50670-901 Recife, PE, Brazil\\
$^{3}$Department of Theoretical Physics, State Polytechnic University, 195251, St.-Petersburg, Russia}

\begin{abstract}
Electromagnetically induced transparency (EIT) has mainly been
modelled for three-level systems. In particular, a considerable
interest has been dedicated to the $\Lambda$-configuration, with
two ground states and one excited state. However, in the
alkali-metal atoms, which are commonly used, hyperfine interaction
in the excited state introduces several levels which
simultaneously participate in the scattering process. When the
Doppler broadening is comparable with the hyperfine splitting in
the upper state, the three-level $\Lambda$ model does not
reproduce the experimental results. Here we theoretically
investigate the EIT in a hot vapor of alkali-metal atoms and
demonstrate that it can be strongly reduced due to the presence of
multiple excited levels. Given this model, we also show that a
well-designed optical pumping enables to significantly recover the
transparency.
\end{abstract}

\maketitle

\section{Introduction}\label{sec1}

Electromagnetically induced transparency (EIT), i.e. the fact that
a strong field resonant with an atomic transition can make the
atomic medium transparent for another field resonant with a
transition sharing the same excited state, has been intensively studied for the last two decades \cite{Harris1990,Boller1991}. It has given rise to
important applications such as Doppler-free spectroscopy,
high-precision magnetometery, and lasing without inversion
\cite{Haris1997}. EIT also led to the demonstration of slow-light
\cite{Hau1999}. This opened the way to the development of a
reversible memory for light based on dynamic EIT, which was first
demonstrated for classical pulses in optically dense atomic media
\cite{Liu2001,Phillips2001}. In the framework of quantum
information processing and networking, which relies critically on
such memories \cite{Kimble,Lvovsky2009,Sangouard2010}, these works have
been extended to the storage of single-photon pulses
\cite{Chaneliere2005, Eisaman2005}. Further developments have
resulted in the
demonstration of reversible mapping of single-photon entanglement
into and out of a quantum memory \cite{Choi2008}. In the regime of
continuous variables, notable advances have been the storage of
squeezed light \cite{Honda2008,Appel2008} and the storage in an
alkali-metal vapor of a faint coherent pulse retrieved without added
excess noise \cite{Cviklinski2008,Ortalo2009}.

The performances of the EIT-based memories are limited by several
sources of losses. Theoretical models usually rely on a
three-level system in a $\Lambda$ configuration:  two atomic
ground states are connected to the same excited state via a
control field on one transition and a signal field on the other
one. In general, these models take into account the losses
occurring when the optical depth is not sufficient for a full
pulse compression in the medium, which leads to some non-zero
transmittance during pulse storage, and also the losses due to
atomic ground state decoherence
\cite{Fleischhauer2005,Hammerer2010}. However, it must be noted
that the experimental demonstrations of such optical and quantum
memories were mostly performed in ensembles of alkali-metal atoms. In
this case, the level structure is more complex than the simple
three-level $\Lambda$ approximation due to the hyperfine
interaction, and the two ground states are often coupled to two,
three or even more excited states by the laser fields. In many
cases, the excited levels are quite close to each other and the
inhomogeneous broadening is comparable with the hyperfine
splitting, such as for example in the $D_2$-line of cesium atoms. This complex structure may
strongly modify the EIT dynamics. Indeed, it has been
experimentally demonstrated that the EIT can
completely disappear when the inhomogeneous broadening is larger
than the hyperfine splitting in the excited state \cite{Li2009}. More generally,
in many experimental studies, the transparency is smaller than the
value predicted by a three-level theoretical model. The aim of
this paper is thus to go beyond the usual three-level $\Lambda$
approximation and investigate in particular the effect of the
inhomogenous broadening in this case.

Several theoretical studies of EIT have taken into account a
double $\Lambda$-system with two excited levels in the context of
a four-wave mixing process \cite{Cerboneschi1996,Korsunsky1999,Glorieux2010},
which leads to lasing without inversion and squeezed light
generation \cite{Fleischhauer2005}. In these models four levels
are considered to be coupled with the fields but each pair of
fields is coupled with only one of the $\Lambda$-channels. Some
investigations have also addressed a different regime where the
two fields are allowed to transfer the atom into a superposition
of two excited states. They have demonstrated inhibition or
enhancement of the off-resonant Raman transition
\cite{Deng1983,Xia1997,Mishina2007,Mishina2010,Sheremet2010,Freegarde2010}. Numerical
analysis of the field transmission through an inhomogeneously
broadened medium of alkali-metal atoms in Ref. \cite{Bhattacharyya2007}
has also shown that several absorption peaks can be observed due
to the velocity selective optical pumping via several excited
states. A shift of the EIT window from the two-photon resonance
and a partial reduction of the transparency due to the presence of
the second excited state were observed and theoretically justified
in Refs. \cite{Hau1999,Chen2009,Deng2001}. These studies show that several effects may deeply modify the
properties of the EIT as compared to the predictions of a simple
$\Lambda$ system.

In this paper, we present a general analysis of EIT addressing
both the case of an inhomogeneously broadened medium and that of
several $\Lambda$ transitions due to multiple excited levels. We
derive a full analytical expression for the atomic susceptibility
that shows evidence for the interference between multiple
$\Lambda$ transitions and unusual velocity dependent light shifts
of the resonances. In a situation for which the Doppler broadening
of the medium is larger than the separation between different
excited states, these effects are responsible for turning the
transparency into absorption for some velocity groups of atoms,
which significantly reduces the EIT peak in such system. This
reduction is commonly observed in EIT experiments on the $D_2$
line of alkaline atoms for vapors at room temperature \cite{Cviklinski2008,Li2009,Lezama1997,Chakrabati2005}. We show that such effect could cause complete disappearance of the EIT peak, if the atoms with reduced transparency were not optically pumped to levels not participating in the EIT process. Once the crucial role
of optical pumping in the system is clarified, we devise a new
optical pumping scheme to significantly enhance the EIT peak in
room temperature atomic vapors. Even though our discussion is
focused on the excitation of alkaline atoms, our model can be applied as well to various
atom-like physical systems presenting large inhomogeneous
broadening and multiple excited levels.

The paper is organized as follows. In Sec. \ref{sec2}, the
theoretical model is presented. Section \ref{sec3} gives the
absorption profile of a single atom with multiple excited states.
We demonstrate that this profile is strongly dependent on the
velocity of the atom. In Sec. \ref{sec4}, we then consider an
ensemble of atoms with different velocities and demonstrate that
the inhomogeneous broadening leads to a drastic decrease of the
transparency. Finally, in Sec. \ref{sec5}, we present a possible
method to enhance the transparency in such configurations by an
effective cooling mechanism based on optical pumping. Section
\ref{sec6} gives the concluding remarks. The details of the
theoretical derivations are presented in Appendices \ref{A} and
\ref{B}.

\section{$\Lambda$-type interaction with multiple excited levels}\label{Theoretical model}\label{sec2}

In this section, we study the influence of the atomic excited
state structure on a $\Lambda$-type interaction between an atom
and two light fields. In order to
investigate the transmission of a  weak probe light through an
atomic medium driven by a strong control field, we first derive here an
analytical expression for the atomic susceptibility.

\subsection{A model case: Cesium $D_2$-line}
A multilevel structure appears in particular in the case of atomic
levels possessing a hyperfine structure. Here we will consider
alkali-metal atoms, which have a non-zero nuclear spin. Optical D-lines
transitions $n^2S_{1/2}\rightarrow n^2P_{1/2}$ ($D_1$-line) and
$n^2S_{1/2}\rightarrow n^2P_{3/2}$ ($D_2$-line) are split due to the
interaction between electron and nuclear spins. In the ground state
$n^2S_{1/2}$ this interaction leads to a hyperfine splitting of
several GHz. The splitting in the excited state is smaller and the
excited hyperfine levels are separated by few hundreds MHz. We will
consider a $\Lambda$ type interaction (Fig.\ref{fig1}) where the two
ground states are sublevels of the same hyperfine state. In view of
the large hyperfine splitting in the ground state, we can neglect
the interaction of the fields with the other hyperfine state. This
is not true in the excited state, where we will have to take into
account an interaction of the fields with several levels of the
hyperfine manifold.

As a specific example, we will consider a cesium atom $^{133}Cs$,
which has been widely used for the experimental investigations of
light-matter interfacing \cite{Lvovsky2009,Hammerer2010}, for high sensitivity
magnetometery \cite{Bedker2007}, as well as for precision
frequency measurements in atomic clocks \cite{Bize2005}. For the
$D_2$ line, the separation between the closest transitions, which
is 150 MHz, is approximately the same as the Doppler linewidth in
an ensemble of cesium atoms around room temperature
(about 300 K). In practice, due to this Doppler broadening, the
transitions are not resolved, and, as we will show, the combined
action of this broadening and of the hyperfine structure will
strongly influence the EIT interaction.

\subsection{Basic assumptions and energy levels}\label{subsec2.1}

\begin{figure}[tpb!]
\vspace{\baselineskip}
\includegraphics[width=0.9\columnwidth]{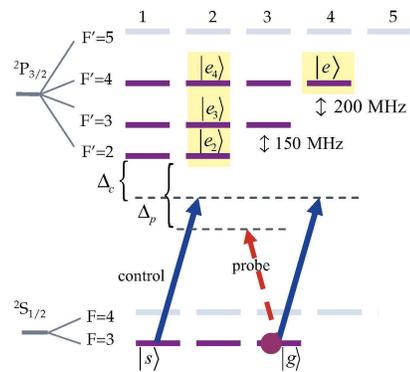}
\caption{(color online) Level scheme of ${}^{133}$Cs $D_2$-line. In
the six-level model, we include the excited levels $|e_2\rangle$,
$|e_3\rangle$, $|e_4\rangle$, $|e\rangle$ and the ground levels
$|s\rangle$ and $|g\rangle$. The three-level approximation
involves the ground levels $|s\rangle$, $|g\rangle$ and only
$|e_2\rangle$ as an excited level.}
\label{fig1}%
\end{figure}%
The scheme of the light-atom interaction in the D$_2$-line of
$^{133}$Cs atoms is sketched in Fig. \ref{fig1}. For a rigorous
study of the multilevel structure influence on the EIT effect we
consider a six-level model, since at least six levels are actually
involved in the interaction. State $|g\rangle\equiv|F=3,
m=3\rangle$ is coupled
with the excited states $|e_2\rangle\equiv|F'=2, m=2\rangle$,
$|e_3\rangle\equiv|F'=3, m=2\rangle$ and $|e_4\rangle\equiv|F'=4,
m=2\rangle$ by a weak $\sigma^-$ polarized probe field. A second
set of atomic transitions is excited by the strong $\sigma^+$
polarized control field coupling the ground state
$|s\rangle\equiv|F=3, m=1\rangle$ with the same excited states
$|e_2\rangle,|e_3\rangle,|e_4\rangle$ as the probe field. The
detunings of the probe with frequency $\omega_p$ and control field
with frequency $\omega_c$ from the atomic transitions
$|g\rangle\rightarrow|e_2\rangle$ and
$|s\rangle\rightarrow|e_2\rangle$ will be denoted respectively
$\Delta_p=\omega_p-\omega_{e_2\,g}$ and
$\Delta_c=\omega_c-\omega_{e_2\,s}$. We will assume the control
field to be close to resonance with the
$|s\rangle\rightarrow|e_2\rangle$ transition. Let us point out that the
dipole moment of the transition $|g\rangle\rightarrow|e_2\rangle$
is the largest of the three transitions
$|g\rangle\rightarrow|e_i\rangle$ accessible to the probe field.

In addition to the $\Lambda$-type interactions, the control field
can couple the states $|g\rangle$ and $|e\rangle\equiv|F'=4,
m=4\rangle$. As the control field is close to
resonance with the $|s\rangle\rightarrow|e_2\rangle$ transition,
this term will be small, but we will see that it can have a non
negligible effect. From this section up to Sec.~\ref{sec5}, we do
not consider the decay of the excited levels to the $F=4$ cesium
ground state, since our aim here is to highlight the role of the
interference between different excitation pathways. In
Sec.~\ref{sec5}, the optical pumping to the $F=4$ ground state
will be taken into account in order to discuss more realistic experimental
situations.

Finally, we will  compare the six-level model to the simple
three-level approximation consisting of the ground states
$|g\rangle$ and $|s\rangle$, and only one excited state
$|e_2\rangle$.

\subsection{Equations for the six-level model}

For the six-level model, the interaction process can be described by the following
Hamiltonian $H$:

\begin{eqnarray}
H =H_{0}+V=H_{\mathrm{field}}+H_{\mathrm{atom}}+V
\end{eqnarray}
with
\begin{eqnarray}
 V&=&V_p+V_c%
 \nonumber  \\
 V_p&=&-\sum_{F^{\prime}=2}^{4}d_{e_{F^{\prime}}\,g}\,|e_{F^{\prime}}\rangle\langle g|E^{(+)}_p + h.c.  \\
 V_c&=&-\sum_{F^{\prime}=2}^{4}d_{e_{F^{\prime}}\,s}\,|e_{F^{\prime}}\rangle\langle s|E^{(+)}_c - d_{eg}\,|e\rangle\langle g|\,E^{(+)}_c + h.c. \nonumber
\label{2.2}
\end{eqnarray}
The Hamiltonian $H_0$ is given by the sum of the free Hamiltonian
operators of the electromagnetic field $H_{\mathrm{field}}$ and of
the atom $H_{\mathrm{atom}}$. The interaction Hamiltonian $V$ is
written in the rotating wave approximation and consists in the
dipole interactions between the atom and the control and probe
fields. In the interaction representation, the positive frequency
components of the electromagnetic field for the control and probe
modes are $E^{(+)}_c=\varepsilon_c\, e^{-i\omega_ct}$ and
$E^{(+)}_p=\varepsilon_p\, e^{-i\omega_pt}$ respectively;
$d_{i\,j}$ is the matrix elements of the electric dipole moment of
the atom between levels $i$ and $j$.

The response of the atom to the weak probe field is described by
the polarizability $\alpha(\Delta_c,\Delta_p)$ defined as
\begin{equation}\label{2.3}
\alpha(\Delta_c,\Delta_p)\,\varepsilon_p= \sum_{F^{\prime}=2}^{4}
d_{g\,e_{F^{\prime}}}\,\sigma_{e_{F^{\prime}}g}\,.
\end{equation}
The right-hand side is written as a function of the steady state
solutions for the slowly varying amplitudes of the optical
coherence between the ground and the excited states addressed by the probe field, which are given by :
\begin{eqnarray}\label{2.4}
  \sigma_{e_{F^{\prime}}\,g}&=&\rho_{e_{F^{\prime}}\,g}\,e^{i\omega_pt} \,,
\end{eqnarray}
where $\rho_{e_{F'}\,g}$ (with $F'=2,\,3,\,4$) are the corresponding
atomic density matrix elements. In the case of a dilute atomic
system, when the number of atoms in a volume of a cubic wavelength
is small, the complex susceptibility is proportional to the single
atom polarizability
\begin{equation}
   \chi(\Delta_c,\Delta_p)=n_0\,\alpha(\Delta_c,\Delta_p)%
   \label{2.5}
\end{equation}
with $n_0$ the atomic density.

The matrix elements of the atomic density matrix $\rho_{ig}$ will
be found in the semiclassical approach as a steady state solution
of the evolution equation up to the first order with respect to
the probe field
$\rho_{e_{F'}g}=\rho^{(0)}_{e_{F'}g}+\rho^{(1)}_{e_{F'}g}$. In the
operator form one needs subsequently to solve the following master
equations
\begin{eqnarray}\label{2.6}
   \frac{d\rho^{(0)}}{dt}&=&\frac{i}{\hbar}\left[\rho^{(0)},H_{\mathrm{atom}}+V_c\right]+\Gamma\left(\rho^{(0)}\right) \,
   \\
   \frac{d\rho^{(1)}}{dt}&=&\frac{i}{\hbar}\left[\rho^{(0)},V_p\right]+%
   \frac{i}{\hbar}\left[\rho^{(1)},\,H_{\mathrm{atom}}+V_c\right]+\Gamma\left(\rho^{(1)}\right)
   \nonumber.
\end{eqnarray}
$\Gamma$ is a relaxation operator describing the radiative decay
of the excited states as well as decoherence processes in the
ground states, as detailed in Appendix \ref{A}.

Here, we assume that the
decoherence in the ground states is much slower than the decay of
the excited states. For instance, in a cesium vapor cell with a
few centimeters in diameter at room temperature, the free flight time
$\tau_d$ of the atom through the beam is on the order of hundreds
of microseconds, which is orders of magnitude larger than the
excited state decay time (tens of nanoseconds). In the following,
we will assume $\tau_d = 300\,\mu$s. In this case the strong
control field optically pumps the atoms in the state with
$m_{F}=3$ for the hyperfine sublevel $F=3$, i.e. in state
$|g\rangle$, and state $|F=3, m_F=1\rangle$, i.e. state
$|s\rangle$ is empty. The zero order density matrix, presented in details in Appendix \ref{A}, has thus only three non-negligible elements $\rho^{(0)}_{g\,g}$,
$\rho^{(0)}_{g\,e}$ and $\rho^{(0)}_{e\,e}$. However, since we
have assumed that the control field is far detuned from the
$|g\rangle\rightarrow|e\rangle$ atomic transition, we can take
$\rho^{(0)}_{g\,g}\approx1$ and
$\rho^{(0)}_{g\,e}\approx\rho^{(0)}_{e\,e}\approx0$. This
approximation is justified for Cesium vapors at room temperature
since the Doppler broadening half width at half maximum is on the
order of 150 MHz and the transition
$|g\rangle\rightarrow|e\rangle$ is separated by 350 MHz from the
resonant transition $|s\rangle\rightarrow|e_2\rangle$, as shown in
Fig. \ref{fig1}. On the other hand, in this part of the
calculation, as mentionned above, we will not take into account
that some atoms decay from the excited states to level
$|F=4\rangle$ in the ground state, and we will assume that the
number of atoms involved in our scheme is kept constant. Optical
pumping towards the $|F=4\rangle$ level and repumping towards
$|F=3\rangle$ will be treated explicitly in section V.

Finally, let us note that in our model the atomic coherence lifetime is only set by the diffusion of the atoms out of the light beam, which corresponds for instance to the case of a cell without paraffine coating or buffer gas. Several studies of the EIT in the presence of atomic diffusion including multiple collisions have been given in \cite{Xiao2006,Kazakov2010,Klein2011}, but in a three-level configuration. In order to focus on the multilevel effect, we do not take such diffusion into account here.

\subsection{Solution for the three-level approximation}

Before solving the six-level model, let us first recall the
solution for a three-level system including the ground states
$|g\rangle$ and $|s\rangle$ and only one excited state
$|e_{2}\rangle$. It is obtained from the previous equations by
setting to zero the
dipole elements with $F'_1,F'_2\neq 2$, $d_{s\,e_{F'_1}}$ and $d_{g\,e_{F'_2}}$. The
coherence then takes the following form
\cite{Fleischhauer2005,Meystre2007}:

\begin{eqnarray}\label{2.7}
\sigma^{(1)}_{e_{2}\,g} &=&
-\frac{\rho^{(0)}_{g\,g}}{2\mathbf{\Delta}_{e_{2}\,g}} \left(1 +
\frac{|\Omega^c_{e_{2}\,s}|^2}{4\mathbf{\Delta}_{s\,g}\mathbf{\Delta}_{e_{2}\,g}}\right)\Omega^p_{e_{2}\,g}.
\end{eqnarray}
The atomic coherence depends on the one- and two-photon detunings
through

\begin{eqnarray}\label{2.8}
\mathbf{\Delta}_{e_{2}\,g} &=& i\gamma_{e_{2}\,g} + \Delta_p
\nonumber
\\
\mathbf{\Delta}_{s\,g} &=& i\gamma_{s\,g} + \Delta_p - \Delta_c -
\frac{|\Omega^c_{s\,e_{2}}|^2}{4\mathbf{\Delta}_{e_{2}\,g}}.
\end{eqnarray}
Here $\Omega^c_{e_{F'}\,s}=2\,d_{e_{F'}\,s}\varepsilon_c/\hbar$
and
$\Omega^p_{e_{F'}\,g}=2\,d_{e_{F'}\,g}\varepsilon_p/\hbar$ (with
$e_{F'}=e_{2}$) are the Rabi frequencies of the control and the
probe fields, respectively. The optical coherence relaxation rate
is $\gamma_{e_{F'}\,g}\approx\gamma/2$ where
$\gamma=2\pi\times5.2$ MHz is the decay rate of the atomic excited
state in the $D_2$-line of $^{133}Cs$ atom and $\gamma_{s\,g}$ is
the decay rate of the ground state coherence $\sigma_{s\,g}$. This
expression gives the well-known dressed atom levels, with a
splitting (Autler-Townes splitting \cite{Autler1955}) of the excited state in two
levels separated by $\Omega^c_{e_{F'}\,s}$ as well as the EIT at $\Delta_p=0$ when $\Delta_c=0$. We
now turn to the multiple level case.

\subsection{Solution for the six-level model}

In the model studied here, which includes four excited states, the
solution for the atomic coherences takes the following form
\begin{eqnarray}\label{2.9}
    \sigma^{(1)}_{e_{F'}\,g}&=&-\frac{\rho^{(0)}_{g\,g}}{2\mathbf{\Delta}_{e_{F'}\,g}}
                  \left(1
                  +\frac{|\Omega^c_{e_{F'}\,s}|^2}{4\mathbf{\Delta}_{s\,g}\mathbf{\Delta}_{e_{F'}\,g}}\right)\Omega^p_{e_{F'}\,g}
                   \nonumber
                  \\
                  &&-\frac{\rho^{(0)}_{g\,g}\Omega^c_{e_{F'}s}}{2\mathbf{\Delta}_{s\,g}\mathbf{\Delta}_{e_{F'}\,g}}
                  \sum_{F'_1\neq F'}
                  \frac{\Omega^c_{s\,e_{F'_1}}}{4\mathbf{\Delta}_{e_{F'_1}\,g}}\Omega^p_{e_{F'_1}\,g}
                  \nonumber
                  \\
                  && -\rho^{(0)}_{g\,g}N_{e_{F'}g}\frac{\varepsilon_p}{\hbar}
\end{eqnarray}
where  $F'$ and $F'_1$ run through the excited states $F',F'_1=2,\,3,\,4$.

The first line coincides with the solution for the three-level
system given in Eq. (\ref{2.8}). The second line comes from  the
presence of several levels in the excited state of the atom, which
introduces additional contributions that can interfere
constructively or destructively with the direct contribution given
in the first line. In the third line, the quantity $N_{e_{F'}g}$
represents the contribution of the excited state $|e\rangle$ which
is small since the control field is far from resonance with the
$|g\rangle$ to $|e\rangle$ transition, as stated above. An
explicit expression for $N_{e_{F'}g}$ is given in the Appendix
\ref{A}.

An important change relative to the three-level model is also the
modification of the detuning term $\mathbf{\Delta}_{s\,g}$ in the
denominator of Eq. (\ref{2.9}), which will generate additional
shifts in the position of the dressed atom levels. The
denominators appearing in Eq. (\ref{2.9}) can now be written as:
\begin{eqnarray}\label{2.10}
&&\mathbf{\Delta}_{s\,g} = i\gamma_{s\,g}+\Delta_p-\Delta_c
            -\sum_{F'}\frac{|\Omega^c_{s\,e_{F'}}|^2}{4\mathbf{\Delta}_{e_{F'}\,g}}
            -\frac{|\Omega^c_{e\,g}|^2}{4\mathbf{\Delta}_{s\,e}}
            \nonumber
            -\mathbf{\Delta}_{N}
    \nonumber
    \\
    &&\mathbf{\Delta}_{e_{F'}\,e} = i\gamma_{e_{F'}\,e}+\Delta_p-\Delta_c-\omega_{e_{F'}\,e} + \omega_{s\,g}
            -\frac{|\Omega^c_{g\,e}|^2}{4\mathbf{\Delta}_{e_{F'}\,g}}
    \nonumber
    \\
    &&\mathbf{\Delta}_{s\,e} = i\gamma_{s\,e}+\Delta_p-2\,\Delta_c+\omega_{e\,e_2}+\omega_{s\,g}
            -\sum_{F'}\frac{|\Omega^c_{s\,e_{F'}}|^2}{4\mathbf{\Delta}_{e_{F'}\,e}}.
    \nonumber
    \\
\end{eqnarray}

Here $\omega_{ij}=(E_i-E_j)/\hbar$ is the atomic transition
frequency between levels $i$ and $j$, and $E_i$ is the energy of
the unperturbed atomic state $|i\rangle$; the optical coherence relaxation rates are
$\gamma_{e\,g}=\gamma_{e\,s}\approx\gamma/2$ and the excited
hyperfine coherence relaxation rate is $\gamma_{e_{F'}\,e}=\gamma$.
The expression of $\mathbf{\Delta}_{N}$ is given in appendix
\ref{A} and similarly to the terms
proportional to $N_{e_{F'}g}$ in Eq. (\ref{2.9}) it will not play a significant role in our analysis.

Substituting expression (\ref{2.9}) into equation (\ref{2.3}) and
using equation (\ref{2.5}) we obtain an analytical expression for
the atomic susceptibility:

\begin{eqnarray}\label{2.11}
\chi(\Delta_c,\Delta_p)&=&-\frac{\rho^{(0)}_{g\,g}n_0}{\hbar}
                         \sum_{F'}\frac{|d_{e_{F'}\,g}|^2}{\mathbf{\Delta}_{e_{F'}\,g}}
                       \nonumber\\
                       &-&\frac{\rho^{(0)}_{g\,g}n_0}{\hbar\,\Delta_{s\,g}}
                         \left(\sum_{F'}
                         \frac{d_{g\,e_{F'}}\Omega^c_{e_{F'}s}}{2\mathbf{\Delta}_{e_{F'}\,g}}
                         \right)^2
                       \nonumber\\
                       &-&\frac{\rho^{(0)}_{g\,g}n_0}{\hbar}\sum_{F'}d_{g\,e_{F'}}N_{e_{F'}g}.
\end{eqnarray}

The expression in the first line corresponds to the sum of the
susceptibilities of independent two-level systems without a
control field. The terms in the second line represent
the effect of multiple $\Lambda$ systems. The squared sum in
parenthesis shows that the contribution of the various $\Lambda$
transitions can interfere positively or destructively. It depends on
the signs of the products of terms such as $d_{g\,e_{F'}}\times
d_{e_{F'}\,s}/\Delta_{e_{F'}\,g}$ and $d_{g\,e_{F"}}\times
d_{e_{F"}\,s}/\Delta_{e_{F"}\,g}$, i.e., the dipole moment between
states $|g\rangle$ and $|e_{F'}\rangle$ coupled by the probe field
and multiplied by the dipole moment between states $|s\rangle$ and
$|e_{F'}\rangle$ coupled by the control field, divided by
$i\gamma$ plus the detuning of the probe field.

To give an example we consider the case of two coupled $\Lambda$
transitions corresponding to two hyperfine upper states, when both
control and probe fields are tuned between the hyperfine levels.
In the situation discussed in this paper the control and the probe
fields have opposite polarizations. In this case the dipole
moments of the neighboring transitions addressed by the probe
field have opposite signs. At the same time, the dipole moments of
the transitions addressed by the control field have the same signs
and the detunings have opposite signs. This leads to a
constructive interference and to an enhancement of the induced
Raman scattering as shown in reference \cite{Mishina2010}.

Let us underline that expression (\ref{2.11}) can be easily
generalized to various physical multilevel systems. Similar
configurations to the hyperfine interaction might appear in
rare-earth doped crystals \cite{Beil2010}, in quantum dots
\cite{Liu2010} or for NV-centers in diamonds \cite{Batalov2009}.

\section{Susceptibility for atoms with non-zero velocity}\label{sec3}

The model has been solved in the previous section for an atom with
zero velocity. However, many EIT experiments are performed in
Doppler broadened media. In order to investigate such
configurations, we first look here at the absorption properties of atoms
with different velocities.

\subsection{Doppler shift}

We consider an atom with velocity $\mathbf{v}$ interacting with
the co-propagating control and probe fields. The Doppler shifts
are approximately the same for the control and probe fields, i.e.
$\Delta_D=-\mathbf{k_c}.\mathbf{v}\approx-\mathbf{k_p}.\mathbf{v}$,
where $\mathbf{k_c}$ and $\mathbf{k_p}$ are respectively the wave
vectors of the control and probe fields. The two-photon detuning
between the control and probe field, which is a critical
parameter, can be thus considered here as independent of the
atomic velocity.

In the following analysis we set the control field detuning from
level  $e_{2}$ equal to zero, $\Delta_c=0$. In this case, the
two-photon resonance will appear when the detuning of the probe
field $\Delta_p$ is also close to zero independently of the atomic
velocity, corresponding to the two-photon resonance. The atomic
absorption profile for different velocity classes is given by the
imaginary part of the susceptibility calculated for the
corresponding Doppler shifts of the two fields
$\mathrm{Im}[\chi(\Delta_p+\Delta_D,\Delta_c+\Delta_D)]$, obtained
from equation (\ref{2.11}).

\subsection{Case of the $\Lambda$ approximation}

\begin{figure}[tpb!]
\vspace{\baselineskip}
\includegraphics[width=0.9\columnwidth]{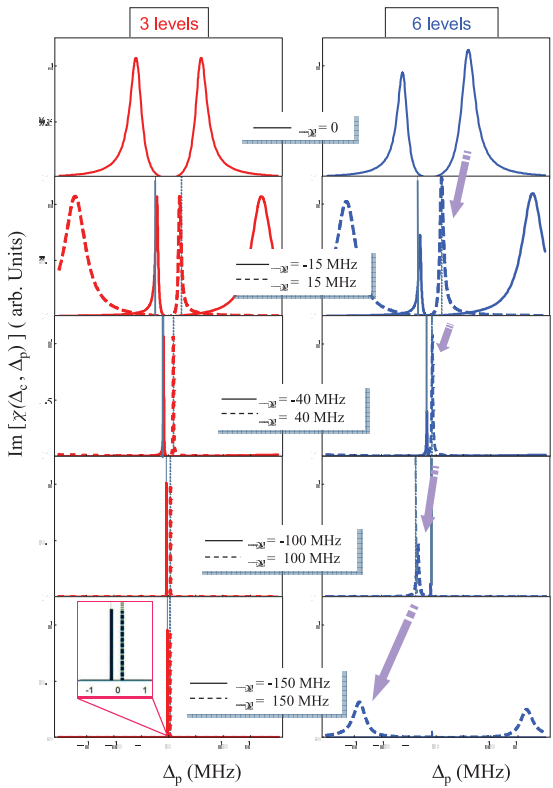}
\caption{(color online) Probe absorption coefficient for atoms with different
velocities as a function of the probe detuning. Calculations for
the three-level and the six-level model are presented respectively
in the first and second column. Dashed curves correspond to the
velocity classes with positive Doppler shifts ($\Delta_D>0$) and
solid curves correspond to the negative Doppler shifts
($\Delta_D<0$) with the values indicated in each graph. Vertical
lines indicate the positions of the resonances approximated by the
analytical equations (\ref{3.2}) and (\ref{3.3}), which are in
good agreement with the precise numerical calculations. In the
three-level system absorption peaks never cross the zero detuning
thus the atomic medium is transparent at $\Delta_p=0$. In
contrast, the full calculation for the six-level system shows that
the induced absorption of atoms from some velocity classes occurs
right at the position where other atoms are transparent. In the
present calculations the control field detuning is $\Delta_c=0$,
the Rabi frequency is $\Omega^c_{e_2s}=2.3\gamma=2\pi\times 12$ MHz and the
ground state decoherence rate is $\gamma_{s\,g}=0.0001\gamma$.}
\label{fig2}%
\end{figure}

Let us first focus on the result for the three-level scheme
presented in the first column of Fig. \ref{fig2}. The positions of
the two absorption peaks  (Autler-Townes doublet
\cite{Autler1955}) are given by the poles of the atomic
susceptibility, which can be obtained by
$\mathrm{Re}[\mathbf{\Delta}_{s\,g}]=0$. We will be interested in
one of the two poles which appears for small probe field detunings
$|\Delta_p|\ll|\Delta_D|$ and is thus located in the EIT window of
atoms with zero Doppler shift. To estimate the position of this
pole, we assume $|\Delta_D|\gg\gamma,\,\,|\Omega_{e_2\,s}|$ and we
rewrite the expression $\mathbf{\Delta}_{s\,g}$ from Eq.
(\ref{2.10}) as:
\begin{eqnarray}\label{3.1}
    \mathbf{\Delta}_{s\,g} &=& i\gamma_{s\,g}+(\Delta_p+\Delta_D)-(\Delta_c+\Delta_D)
              \nonumber\\
              &&-\frac{|\Omega^c_{s\,e_2}|^2}{4(i\gamma_{e_2\,g}+(\Delta_p+\Delta_D))}
              \\
              &\approx&i\gamma_{s\,g}+\Delta_p
              -\frac{|\Omega^c_{s\,e_2}|^2}{4(i\gamma_{e_2\,g}+\Delta_D)}
              \nonumber\\
              &\approx&i\left(\gamma_{s\,g}
              +\frac{\gamma}{2}\frac{|\Omega^c_{s\,e_2}|^2}{4\Delta^2_D}\right)
              +\Delta_p-\frac{|\Omega^c_{s\,e_2}|^2}{4\Delta_D}.\nonumber
\end{eqnarray}

We find that the Autler-Townes absorption resonance (ATR) appears
when the probe field detuning is
$\Delta_p\approx\Delta^{3-level}_{ATR}$ with
\begin{equation}\label{3.2}
    \Delta_p\approx\Delta^{3-level}_{ATR}=\frac{|\Omega^c_{s\,e_2}|^2}{4\Delta_D}.
\end{equation}
This position of the induced absorbtion resonance for the probe
field is thus determined by the dynamic Stark shift of the atomic
levels due to the off-resonant interaction with the strong control
field. The result of the exact calculation for the absorption
spectrum of three level atoms is shown in Fig. \ref{fig2}, first
column. The vertical lines indicate the positions of the resonances
approximated by the analytical equations (\ref{3.2}). The shift is
positive (blue shift) for atoms traveling in a direction opposite
to the lasers ($\Delta_D>0$), while it is negative (red shift) for
atoms traveling in the same direction as the lasers
($\Delta_D<0$). $\Delta^{3-level}_{ATR}$ can be very small but it
must be stressed that it never reaches zero, i.e. the absorption
resonance never takes place at zero detuning. Whatever their
velocity, the atoms are transparent for the probe field at
$\Delta_p=0$.

\subsection{Case of the full model}

As shown in the second column of Fig. \ref{fig2}, the situation is
significantly different for the six-level model. The positions of
the induced Raman absorption resonances for the probe field are
strongly modified. They can be estimated similarly to the
three-level case as a real part of the pole of the atomic
susceptibility closest to the control laser frequency. By setting
to zero the real part of the detuning $\mathbf{\Delta}_{s\,g}$
from equation (\ref{2.10}), we find that the absorption resonance
appears when the probe field detuning is
$\Delta_p\approx\Delta^{6-level}_{ATR}$ with
\begin{eqnarray}\label{3.3}
    \Delta^{6-level}_{ATR}&=&\Delta^{3-level}_{ATR}
                       +\frac{|\Omega^c_{e_3\,s}|^2}{4(\Delta_D-\omega_{e_3\,e_2})}
                       +\frac{|\Omega^c_{e_4\,s}|^2}{4(\Delta_D-\omega_{e_4\,e_2})}
                       \nonumber\\
                       &-&\frac{|\Omega^c_{eg}|^2}{4(\Delta_D-\omega_{e\,e_2}-\omega_{sg})}.
\end{eqnarray}
To derive this expression we assumed that none of the transitions
are saturated by the off-resonant control field:
\begin{eqnarray}\label{3.4}
   \Delta_D&&\gg\gamma,\,\Omega^c_{e_2\,s}%
   \nonumber\\
   \Delta_D-\omega_{e_3\,e_2}&&\gg\gamma,\,\Omega^c_{e3
   \,s}%
   \nonumber\\
   \Delta_D-\omega_{e_4\,e_2}&&\gg\gamma,\,\Omega^c_{e_4\,s}.
\end{eqnarray}
As explicitly written, the first term in the expression
(\ref{3.3}) coincides with the three level approximation
(\ref{3.2}). The second and the third terms show the dynamic Stark
shifts due to the presence of the extra excited states
$|e_3\rangle$ and $|e_4\rangle$. The last term represents the
shift of the state $|g\rangle$ due to the off-resonant action of
the control field on the $|g\rangle\leftrightarrow|e\rangle$
transition. In the second column of Fig. \ref{fig2} we compare
this estimation, represented by the grey vertical lines, with the
exact absorption spectrum calculated numerically for Doppler
shifts equal to 20 MHz, 50 MHz and 100 MHz. As can be seen, they
coincide very well, which confirms the validity of the
approximation made in equation (\ref{3.3}).

It can be clearly seen in Fig. \ref{fig2} (second column) that the
positions of the induced absorption resonances in the six-level
system do not converge to the same point for positive and negative
Doppler shifts as they do in the three-level model. For atoms
traveling in the same direction as the lasers ($\Delta_D<0$, full
lines), the shift of the Autler-Townes absorption resonance always
keeps the same negative sign, since $\omega_{e_3\,e_2}$ and
$\omega_{e_4\,e_2}$ are positive. On the contrary, for atoms
traveling opposite to the light beams ($\Delta_D>0$, dashed
lines), the position of the atomic dressed state moves
continuously from positive to negative detunings when the Doppler
shift increases. For $\Delta_D=150$MHz (right column, bottom
curve), the lasers are resonant with level $e_3$ giving rise to a
splitting in two symmetrical dressed levels.
Because of this dependence of the absorption resonance positions
with respect to velocity, the response of an atomic vapor
containing atoms with all these velocity classes is strongly
modified.

\begin{figure}[tpb!]
\vspace{\baselineskip}
\includegraphics[width=0.9\columnwidth]{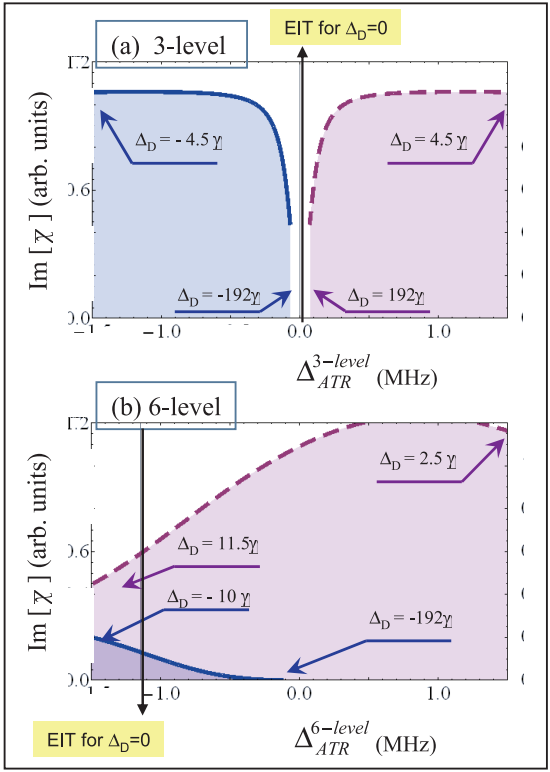}
\caption{(color online) Maximum absorption coefficient versus
position $\Delta_{ATR}$ for different velocity classes with
Doppler shift $\Delta_D$. (a) For the three-level model, the Doppler shift is varied from $-192\gamma$ ($-1000$ MHz) to $-4.5\gamma$ ($-23$ MHz) (blue, solid line) and from $4.5\gamma$ to $192\gamma$ (red, dashed line). $\Delta^{3-level}_{ATR}$ is an exact solution of  $\mathrm{Re}[\mathbf{\Delta}_{s\,g}]=0$ for the three-level system. (b) For the six-level model, the Doppler
shift is varied from $-192\gamma$ to $-10\gamma$ ($-52$ MHz) (blue, solid line) and
from $2.5\gamma$ ($13$ MHz) to $11.5\gamma$ ($60$ MHz) (red, dashed line). $\Delta^{6-level}_{ATR}$ is given by Eq. (\ref{3.3}). The control field is on
resonance with level $|e_{2}\rangle$, $\Delta_c=0$, and the Rabi frequency
is $\Omega^c_{e1\,s}=2\pi\times 12$ MHz. Vertical lines show the positions of
the transparency window for the atoms with zero Doppler shift,
which is at $\Delta_p=0$ for the 3-level system and which is at $\Delta_p=-1.15$ MHz for the six-level system according to Eq.~(\ref{4.3}).}
\label{fig3}%
\end{figure}%

In Fig. \ref{fig3}, we have plotted the heights of the
induced absorption resonances as a function of their positions.
The corresponding Doppler shifts are indicated on the curves. Let
us first examine the three-level model (Fig. \ref{fig3}(a)).
Scanning the Doppler shift from -23 MHz to -1000 MHz, we can see
that the position of the absorption resonance approaches zero for
large Doppler shift. Simultaneously the maximum absorption
decreases due to the dephasing in the ground state. This dephasing
comes from the ground state decoherence rate $\gamma_{s\,g}$ due
to the finite size of the light beam. We set the decoherence to be
equal to the inverse atomic time of flight through the light beam
$\gamma_{s\,g}=0.0001\gamma$. For positive Doppler shift the
result is completely symmetric relative to the zero point. There
is no velocity class producing an induced absorption for
$\Delta_p=0$ and thus the transparency window remains open even if
atoms with all possible Doppler shifts interact with the fields at
the same time.

The six-level model is presented in Fig. \ref{fig3}(b). For
negative Doppler shifts (blue, solid line), due to the
 interaction of several $\Lambda$-channels, the cross
section of the induced Raman scattering is reduced in the wing of
the $D_2$ line. This can be seen from Fig. \ref{fig2} where the
induced absorption resonance almost disappears for $\Delta_D\leq-100$
MHz. This effect was described in reference
\cite{Xia1997,Mishina2010} for the $D_1$-transition in
alkali-metal atoms. On the other hand, when the Doppler shift is
positive (red, dashed line), the situation changes completely. As could already be
seen in Fig. \ref{fig2}, absorption from atoms with non zero Doppler shifts appears for the same probe
field detuning as EIT for atoms with zero Doppler shift. The
position of the EIT for zero velocity is indicated in Fig.
\ref{fig3}(b) by the vertical grey line. We see that the EIT can
be strongly reduced by the combined effect of different
velocity classes.

To summarize our results, this section shows that it is a unique
property of the three-level system that independently of the
atomic velocity the EIT appears at the same probe field detuning.
This allows for the observation of the EIT even in the presence of
a large Doppler broadening \cite{Boller1991}. As demonstrated
here, this property is not preserved in a system with more than
one excited state. There is no longer a probe field detuning for
which atoms with different velocities are transparent. This will
strongly affect the EIT observation in the six-level system in the
presence of a Doppler broadening which is comparable to the
separation between the excited states. This situation is described
in the following section.

\section{Effect of Doppler broadening on EIT for a system with multiple excited levels}\label{sec4}

At room temperature the Doppler broadening for alkali-metal atoms is
comparable with the hyperfine separation between the excited
states of the $D$-transition, as shown in Table \ref{T1}
\cite{Steck2009}. This section studies the influence of a finite
temperature on the probe transmission in a multiple level
configuration.

\begin{table}[htpb!]
\caption{Doppler broadening and hyperfine splittings for cesium
and rubidium atoms at $T=300$K. The Doppler width is given by
$\Gamma_D=\sqrt{k_{B}T/m\lambda^2}$ where $k_B$ is a Boltzmann constant, $m$ is is an atomic mass and $\lambda$ is a light wavelength.}\vspace{5mm}
\begin{tabular}{c c c c c}
 \hline \hline
                &   &            &           &              \\
 Atom  &  &\,\,\, $^{133}Cs$ \,\,\,\,\,\,\, &\,\,\,  $^{87}Rb \,\,\,\,\,\,\,$ &\,\,\, $^{85}Rb$\,\,\,\,\,\,\, \\
                         &   &            &           &              \\ \hline
 Doppler                 &   &             &           &              \\
 width,                  &   & 160        & 198       & 200          \\
     MHz                 &   &            &           &              \\ \hline
                         &   &            &           &              \\
 \,\,\,Hyperfine\,\,\,   &$D_1$-line \,\, \vline   &  1168& 817 &  362   \\
 splitting,              &                          &      &     &        \\
 MHz           &$D_2$-line \,\, \vline  & 151   & 72  &  29   \\
               & \,\,\,\,\,\,\,\,\,\,\,\,\,\,\,\,\,\,\,\,\,\,\, \vline          & 201   &  157&  63   \\
                         & \,\,\,\,\,\,\,\,\,\,\,\,\,\,\,\,\,\,\,\,\,\,\, \vline &  251 &  267&  120  \\
                         &                          &      &     &   \\ \hline \hline
\end{tabular}
\label{T1}
\end{table}
\subsection{Susceptibility of the medium}
\begin{figure}[tpb!]
\vspace{\baselineskip}
\includegraphics[width=0.9\columnwidth]{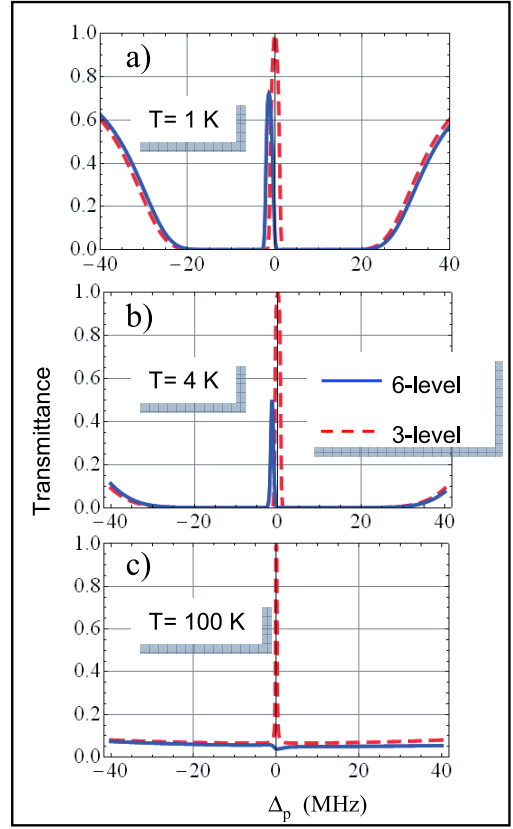}
\caption{(color online) Probe field transmittance $t$ for a Doppler
broadened $D_2$-line for ${}^{133}$Cs vapor with equal number of
atoms and different temperatures: (a) $T$=1 K, $\Gamma_D$= 10 MHz,
(b) $T$=4 K, $\Gamma_D$=20 MHz and (c) $T$=100 K, $\Gamma_D$=100
MHz. The control field detuning is $\Delta_c=0$ and the Rabi
frequency $\Omega^c_{e1s}=2.3\gamma= 2\pi\times 12$ MHz. Calculations are
done with the three-level model (dashed, red curves) and the
six-level model (solid, blue curves). The atomic density is the
same for curves a, b and c and is equal to $n_0=1.1\times10^{10}$
cm$^{-3}$ with all atoms in state $|g\rangle$.}
\label{fig4}%
\end{figure}

To find the susceptibility of the sample at a given temperature we
average the susceptibility of the atoms
$\chi(\Delta_c+\Delta_D,\,\Delta_p+\Delta_D)$ over the velocity
distribution $f(\Delta_{D})$ for atoms in the ground state
$|g\rangle$:
\begin{equation}
    \bar{\chi}(\Delta_c,\,\Delta_p)=\int\chi(\Delta_c+\Delta_{D},\Delta_p+\Delta_{D})f(\Delta_{D})d\Delta_{D}.%
    \label{4.1}
\end{equation}
The probe transmittance
$t=|\varepsilon_{p}(L)|^2/|\varepsilon_{p}(0)|^2$ through the
medium of length $L$ is given by the Beer's law:
\begin{eqnarray}
 t=\exp\left(-4\,\pi\,k_p\,L\,\mathrm{Im}[\bar{\chi}(\Delta_c,\Delta_p)]\right).
 \label{4.2}
 \end{eqnarray}
The velocity distribution is assumed to be Gaussian with
$f(\Delta_D)=(2\pi\Gamma^2_D)^{-1/2}\exp(-\Delta^2_D/2\Gamma^2_D)$
where the Doppler width $\Gamma_D$ depends on the temperature $T$
of the sample.
In Fig. \ref{fig4} we present the probe transmittance $t$ as a
function of the probe field detuning $\Delta_p$ for a cesium vapor
with different temperatures. For comparison, the figure gives the
results for the three-level and the six-level models.

\subsection{Case of small Doppler broadening}
Let us consider first the case of a sample with small Doppler
broadening. When the temperature of the medium is $T=1$ K and the
Doppler broadening is $\Gamma_D=10$ MHz, which is much smaller
than the splitting between the closest hyperfine excited states, a
clear EIT resonance is predicted by both models. It is indeed
well-known that the transparency is present in cold cesium atoms
despite the complicated multi-level structure \cite{Choi2008}.
However, there are some noticeable differences in the predictions
of the two models (see Fig. \ref{fig4}).

First, the EIT resonance is shifted from the bare two photon
resonance $\Delta_p=0$. This light shift is caused by he presence of
the excited states $|e_3\rangle,\,|e_4\rangle$ and $|e\rangle$, as
previously explained in references \cite{Hau1999,Mishina2010} in the
case of  the $D_1$-transition in alkali-metal atoms. In the presence
of additional excited states, there are additional dynamic Stark
shifts of the dressed states due to the off-resonant interaction
with the control field. The EIT
resonance is located at the minimum of $\mathrm{Im}[\sigma^{(1)}_{e_{2}\,g}]$
given by Eq.~\ref{2.9}. If the separations between the hyperfine
transitions are larger than the Rabi frequencies of the control
field $\omega_{e_3\,e_2}\gg\Omega^c_{e_3\,s}$ and
$\omega_{e_4\,e_2}\gg\Omega^c_{e_4\,s}$ the EIT position can be approximated by the following expression:
\begin{equation}\label{4.3}
     \Delta^{6-level}_{EIT}=-\frac{|\Omega^c_{e_3\,s}|^2}{4\omega_{e_3\,e_2}}
                           -\frac{|\Omega^c_{e_4\,s}|^2}{4\omega_{e_4\,e_2}}
                       +\frac{|\Omega^c_{eg}|^2}{4(\omega_{e\,e_2}+\omega_{sg})}.
\end{equation}
The  first two terms in expression (\ref{4.3}) represent the
shifts of the EIT point due to off-resonance excitations of the
states $|e_3\rangle$ and $|e_4\rangle$. The last term corresponds
to the shift of the state $|g\rangle$ due to the action of the
off-resonant control field on the
$|g\rangle\leftrightarrow|e\rangle$ transition.

Second, the peak transmittance is reduced in a multilevel
configuration. This effect can be explained by the dephasing of
the ground state coherence $\sigma_{s\,g}$ introduced by the
off-resonance excitation of the levels $F'=3$ and $F'=4$
by the control field. Both effects were observed in Ref.
\cite{Hau1999} and studied in details in Refs.
\cite{Chen2009,Mishina2010}.

\subsection{Case of broadening of the order of the hyperfine splitting}

If we increase the medium temperature such that the Doppler
broadening becomes comparable with the hyperfine splitting in the
excited state, the peak transmittance drops down and finally
disappears as shown in Fig. \ref{fig4}b and \ref{fig4}c. At the
same time the three-level model predicts only a narrowing of the
transparency window in agreement with Ref. \cite{Figueroa2006}. This
modification can be attributed to the six-level atoms which
have Doppler shifts in the interval $\Delta_D\in[20, \,
100]$ MHz. Some of them are not transparent for the probe field as
we concluded from the previous section. From Fig. \ref{fig3}, it
can be seen that the atoms moving in a direction opposite to the
laser beam ($\Delta_D>0$) are at the origin of this effect.

\begin{figure}[tpb!]
\vspace{\baselineskip}
\includegraphics[angle=270,width=0.99\columnwidth]{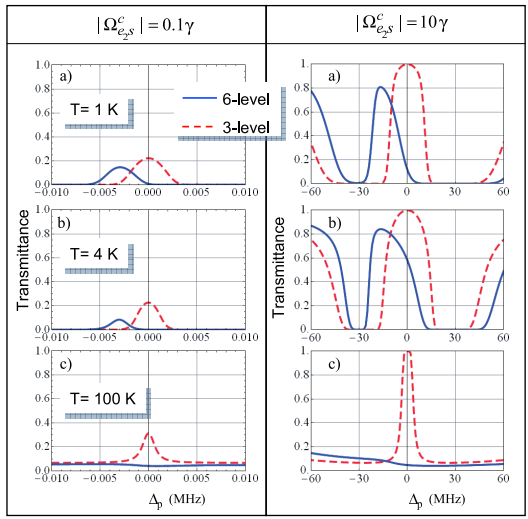}
\caption{(color online) Probe field transmittance $t$ for a Doppler
broadened $D_2$-line for ${}^{133}$Cs vapor for different control Rabi frequencies. The parameters and notations are similar than in figure \ref{fig4}. The Rabi
frequency are here (left column) $\Omega^c_{e1s}=0.1\gamma= 2\pi\times 0.52$ MHz and (right column) $\Omega^c_{e1s}=10\gamma= 2\pi\times 52$ MHz. In both cases, large inhomogeneous broadening leads to the total disappearance of the transparency peak. Let us note that the reduced transmittance in the three-level model for small Rabi frequency is due to a reduced ratio between this frequency and the decoherence rate taken equal to $\gamma_{sg}=0.0001\gamma$ in all the simulations.}
\label{fig5}%
\end{figure}

Let us note that our study has focused on a regime where the control Rabi frequency is on the order of the excited state linewidth $|\Omega^c_{e_2s}|\sim\gamma$. However, it can be shown that the strong reduction of the transparency will occur for smaller or larger frequencies. Figure \ref{fig5} gives the transmittance of the medium for $|\Omega^c_{e_2s}|=0.1\gamma$ and $|\Omega^c_{e_2s}|=10\gamma$. In both cases, large inhomogeneous broadening leads to the total disappearance of the transparency peak.

Experimentally, an EIT peak is still observed even at room
temperature, but it is strongly reduced. As we demonstrate in the
next section, the non-zero value of the EIT signal, even if
considerably degraded, is caused by some optical pumping to $F=4$
of the atoms that are not transparent to the probe field. This
process was not taken into account in our theoretical model up to
this point, but it will be treated in the next section.

This section demonstrated then that the EIT is strongly reduced in
an inhomogeneously broadened ensemble of six-level atoms as
compared to three-level atoms. The main reason has been clearly
identified: it is the presence of the atoms with a particular
Doppler shift that absorb the probe light while the others are
nearly transparent. Based on this result, the following section
investigates a scheme possibly enabling the improvement of the EIT
in such systems.

\section{EIT enhancement by velocity selective optical pumping}\label{sec5}

The velocity classes of atoms leading to a strong reduction of the
EIT have been clearly identified in the previous sections. In
order to recover the transparency, these atoms should not
participate in the process.  In this section we discuss
successively two optical pumping schemes that modify the velocity
distribution in order to enhance the transparency.

\subsection{Optical pumping by the strong control field}\label{subsec5.1}

Up to now we have considered a six-level system as a close
approximation to the $D_2$-line of the cesium atom (Fig.
\ref{fig1}). We have taken into account only one ground hyperfine
sublevel $F=3$ which was coupled with the excited $6^2P_{3/2}$
manifold by both control and probe fields. In this case the strong
$\sigma^+$ polarized control field optically pumps the atoms into
state $|g\rangle$. However, during the optical pumping process the
atom can spontaneously decay from the excited sates to both ground
hyperfine sublevels $F=3$ and $F=4$. Taking the second hyperfine
sublevel in the ground state into account leads to the fact that
all atoms are eventually pumped to the $F=4$ state. Even atoms
firstly pumped into state $|g\rangle$ can be re-pumped via state
$|e\rangle$ to the upper ground state with $F=4$. To bring back
atoms in state $|g\rangle$ a $\sigma^+$ repumping  polarized field
resonant with the $F=4 \rightarrow F'=4$ transition is applied.
The repumping field (together with the control field), can bring a
non-zero steady state population back into state $|g\rangle$.

Furthermore, the optical pumping process depends on the atomic
velocity and can strongly modify the velocity distribution. To
quantitatively estimate this effect we have simulated the entire
optical pumping process and obtained a numerical steady state
solution of the equations for the atomic density matrix using the
following interaction Hamiltonian:
\begin{eqnarray}
V_{OP}=&-&\sum^4_{F=3}\sum^{F+1}_{F'=F-1}\sum^{F}_{n=-F}\hbar\Omega_{F'\,n+1\,\,F\,n}|F',n+1\rangle\langle
F,n|
          \nonumber
          \\
          &+&h.c.%
\label{5.1}
\end{eqnarray}
Here the Rabi frequency $\Omega_{F'\,n+1\,\,F\,n}$ for the
transition $|F,n\rangle \rightarrow |F',n+1\rangle$ contains the
control field amplitude if $F=3$ and the repump field amplitude if
$F=4$. Due to the complexity of the atomic level structure the
full set of equations for the atomic density matrix is presented
in Appendix \ref{B}. Here we comment the main features of our
optical pumping model and the results.

The efficiency of the optical pumping is limited by the relaxation
processes. During the collisions with the cell walls the atomic
spin as well as the velocity is changing. To model this process we
introduce in our theory the decay of the ground state density
matrix elements. The decay rate $1/\tau_d$ is assumed to be much
smaller than the excited state decay rate $\gamma$, so we can
neglect the influence of the wall-collisions on the atom in the
excited state. In our model, we assume that the randomisation
times for the atomic velocity and for the atomic spin are equal,
which can describe at least three experimental situations. First,
when the cell walls do not have special coating preventing the
spin relaxation and after one wall collision both atomic spin and
velocity are randomly changed. Second, when the cell has a
polarization preserving coating but the volume of the light beam
is much smaller than the volume of the cell. In this case an atom
will collide many times with the wall and thus randomly change its
spin before coming back to the light beam. A third situation
occurs when the optical pumping process reaches its steady state
during the time it takes for an atom to cross the beam. In these
cases, in order to find the steady state, we set $\tau_d$ equal to
the time of flight of the atom through the light beam.

\begin{figure}[tpb!]
\vspace{\baselineskip}
\includegraphics[angle=270,width=0.9\columnwidth]{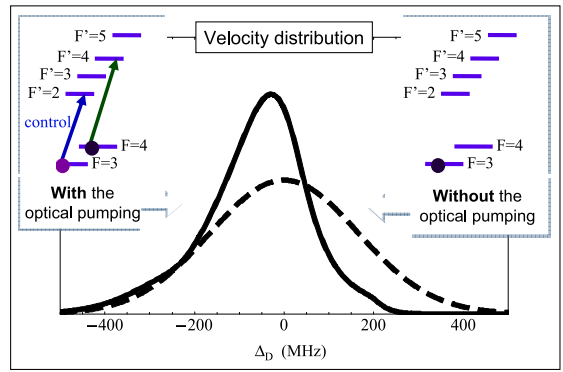}
\caption{(color online) Velocity distribution of the atoms in
state $|g\rangle$. The distributions are normalized such as $\int
f(\Delta_D)d\Delta_D=1$. The dashed line represents the Gaussian
distribution, assuming that all the atoms are prepared in state
$|g\rangle$ as shown in the right inset. The solid line shows the
actual velocity distribution which can be achieved with the
transfer of atoms due to optical pumping by the control field and
the repumping field as shown in the left inset. The parameters are
the following: $\Omega^c_{e1s}=2.3\gamma=2\pi\times 12$ MHz, $\Delta_c=0$, repumping power equal to $4.4$ mW for a beam diameter of $1$ cm, $\tau_D=300 \mu s$.}
\label{fig6}%
\end{figure}%

In our calculations, $\tau_d$ is chosen equal to 300$\mu s$, which
corresponds to an average time of flight of several centimeters for
the cesium atoms at room temperature. The density of atoms was set
to $n_0=3.5\times10^{11}$ cm$^{-3}$. This number is an order of
magnitude larger than the one we used in the previous section,
because here we consider the two ground states $F=3$ and $F=4$.

We will keep the Rabi frequency of the control field to be the
same as in the previous sections, $\Omega^c_{e_2\,s}=2\pi\times 12$ MHz. This
corresponds, for example, to a control field power of $200$ mW and
a beam diameter equal to $1$ cm. The repumping field is
contra-propagating with respect to the control and probe fields,
and its power is equal to $4.4$ mW for a beam diameter of $1$ cm.

Figure \ref{fig6} gives the result of the optical pumping
simulation. The velocity distribution of the atoms in state
$|g\rangle$ in the presence of the control and repumping beams is
shown by the solid line. For comparison we also give the Gaussian
distribution with a width $\Gamma_D=160$ MHz corresponding to room
temperature $T=300$ K. We can see that the atoms propagating
towards the control beam ($\Delta_D>0$) have been pumped out of
state $|g\rangle$. This is due to the positive Doppler shift that
brings the control light frequency closer to the atomic transition
$|F=3\rangle \rightarrow |F'=4\rangle$ and thus makes depopulation
of the state $|g\rangle$ more likely. As a consequence, the
velocity distribution of atoms in state $|g\rangle$ becomes
narrower, indicating an effective cooling mechanism.

\begin{figure}[tpb!]
\vspace{\baselineskip}
\includegraphics[angle=270,width=0.85\linewidth]{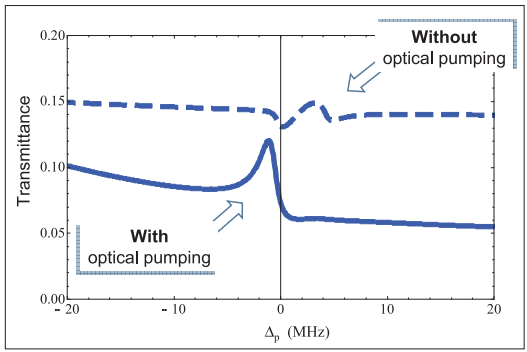}
\caption{(color online) Transmittance for the probe field with and without
optical pumping. The dashed curve corresponds to six-level atoms
in state $|g\rangle$ having a Gaussian velocity distribution. The
EIT peak vanishes in these conditions as shown in the previous
section. The solid line shows the transmittance
 when the velocity distribution is modified by the
optical pumping due to the control and the repumping fields. The
EIT peak is partly recovered since part of the absorbing atoms are
pumped out to sublevel $F=4$ and thus do not interact any more
with the probe field. The parameters are the same as for Fig.~\ref{fig6}}
\label{fig7}%
\end{figure}%

According to the conclusion of section \ref{sec3}, atoms with
positive Doppler shifts are the ones which previously contributed
to the disappearance of the EIT. Due to the optical pumping
process, these atoms, which move in the opposite direction to the
probe and control beams, are partially removed from the
interaction process.
 Using the new velocity distribution in the calculations we can see
how it helps to recover the transparency. In Fig. \ref{fig7}, we
show the transmittance of a cesium vapor calculated including the
optical pumping effect. Contrary to the case of a Gaussian
velocity distribution, in which the EIT resonance vanishes, a
transparency peak is obtained with the modified velocity
distribution. In \cite{Cviklinski2008}, EIT was indeed
experimentally observed on the $D_2$ line of Cs, while the combined
effect of broadening and multi-level structure should lead to a
fully vanishing of the EIT.

\begin{figure*}[htpb!]
\vspace{\baselineskip}
\includegraphics[angle=270,width=1.85\columnwidth]{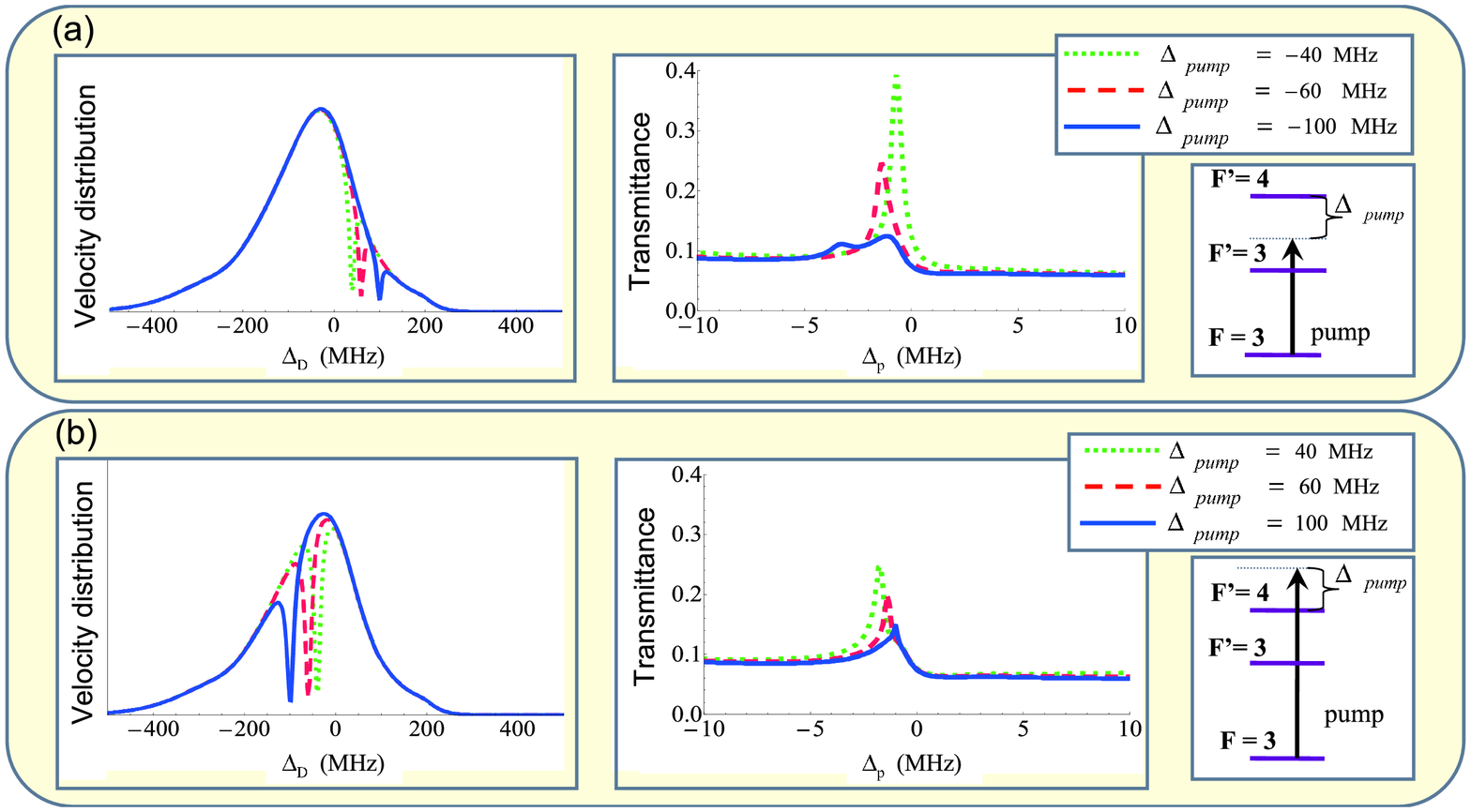}
\caption{(color online) Influence of the additional pump on the
velocity distribution of atoms in state $|g\rangle$ (1st column)
and on the probe field transparency (2nd column). The optical
pumping leads to a hole burning effect in the velocity
distribution, which results in an enhancement of the EIT.
Different probe field detunings are presented with
$\Delta_{pump}<0$ in the upper panel and $\Delta_{pump}>0$ in the
lower panel. The Rabi frequency of the pump field with respect to
the transition between $|g\rangle$ and $|e\rangle$ sublevels
remains constant $\Omega_{pump}=0.15\gamma$. Calculations are done
in the same conditions for the control and repumping fields as in
Fig. \ref{fig6}: $\Omega^c_{e1s}=2.3\gamma=2\pi\times 12$ MHz, $\Delta_c=0$,
repumping power equal to $4.4$ mW for a beam diameter of $1$ cm,
$\tau_D=300 \mu s$.}
\label{fig8}%
\end{figure*}%

We have thus demonstrated that the effective cooling due to the
optical pumping process by the control field allows to observe the
EIT effect in the configuration presented in Fig. \ref{fig1}.

\subsection{Hole burning in the velocity distribution}\label{subsec4.2}

Based on the previous result, we now present a procedure that allows
to further enhance the EIT contrast without changing the medium
optical depth. The main idea is to burn a hole in the velocity
distribution of atoms in state $|g\rangle$. Thus we exclude atoms
with a specific Doppler shift from the interaction process.

In addition to the control, probe and repumping fields discussed
in the previous subsection, one more $\sigma^+$ polarized pump
field is used. It is detuned by $\Delta_{pump}$ from the
transition between $F=3$ and $F'=4$ sublevels. It will then pump
the atoms with a Doppler shift $\Delta_D=-\Delta_{pump}$ out of
the $F=3$ level, thus creating a hole in the velocity
distribution. We simulate this additional optical pumping process
by introducing an effective depopulation rate $\gamma_{pump}$ for
the atoms in state $|F=3,m_F=3\rangle$:
\begin{eqnarray}\label{5.2}
    \gamma_{pump}(\Delta_D)&=&\frac{\gamma\,\Omega_{pump}^2}
                               {4(\Delta_D+\Delta_{pump})^2+\gamma^2+\Omega_{pump}^2}
\end{eqnarray}
where $\Omega_{pump}$ is the Rabi frequency of the pump field with
respect to the transition $|g\rangle\leftrightarrow|e\rangle$. The
structure of this expression shows a Lorentzian dependence with
the Doppler shift. The width of this Lorentzian curve, which gives
the width of the dip on the velocity distribution, depends in the
Rabi frequency of the field but it is never smaller than the
natural linewidth of the atomic excited state $\gamma$.

Figure \ref{fig8} shows the velocity distribution of atoms in
state $|g\rangle$ and the resulting transmittance for the probe. We examine
the cases of different values of the pump field detuning
$\Delta_{pump}$, keeping the pump Rabi frequency equal to
 $\Omega_{pump}=0.15\gamma$. Comparing the cases of negative and
positive detunings, we see that the EIT enhancement is larger when
the pump detuning is negative. This is due to the fact that atoms
with positive Doppler shifts, pumped out of the interaction process in this case, contribute more to the
absorption than atoms with negative Doppler shifts as shown in Fig. \ref{fig3}(b). We see that the EIT
enhancement is the largest when $\Delta_{pump}=-40$ MHz. We should
also note that that the EIT contrast grows when the pump Rabi
frequency is increased because more absorbing atoms are depumped
and thus removed from the interaction process.

This result confirms our interpretation of the EIT formation in a
Doppler broadened $D$-line of the alkali-metal atoms. One can define the
EIT contrast as $C\equiv ~\!(~\!\mathrm{t}_{max}~-~\mathrm{t}_{min}~)\!~/\!~(\!~1~-~\mathrm{t}_{min}~)$, where
$t_{max}$ is the probe transmittance in the maximum of the EIT peak
and $t_{min}$ is the probe transmittance on the plateau, which
defines the optical depth of the medium on the side of the EIT
resonance. It can be seen in Fig. \ref{fig8} that our scheme
provides an enhancement of this contrast by approximately $8$ times.

\section{Conclusion}\label{sec6}

Our analysis demonstrates the influence of multiple excited levels
on EIT in alkali-metal atoms. We have shown that the presence of
more that one excited state in the $\Lambda$-type interaction can
lead to a significant change as compared with the three-level
approximation. This change is small when the inhomogeneous
broadening is much smaller than the separation between the excited
states. However, even for cold atoms, it leads to a noticeable
decrease in transparency.

The effect becomes large when the broadening is comparable with
the separation between the excited states, which may lead to the
total disappearance of the transparency. This effect of the
multilevel structure is caused by two reasons. The first one is
the ac-Stark shift of the atomic dressed states, which varies with
the Doppler shift. In a multilevel atom this variation leads to
having no probe-field frequency for which atoms from all the
velocity classes are transparent. The second reason is the
interference between several $\Lambda$ transitions appearing when
both fields are tuned between the hyperfine transitions,
increasing the off-resonant Raman scattering cross section.

After explaining the origin of the EIT suppression in a Doppler
broadened medium, in particular by clearly identifying the atoms
which strongly absorb the probe, we have proposed a method to
enhance the transparency. It is based on an effective cooling
mechanism using optical pumping. In particular, by creating a dip in the velocity
distribution of atoms participating in the interaction process we
achieve a significant enhancement of the EIT contrast. This
technique is expected to allow improvements in current
experimental realizations of EIT-based quantum memories in room
temperature alkali-metal vapors. Our analysis might also be
relevant for other coherent effects in such vapors, like in the
recent entangled light generation by four-wave mixing
\cite{Lett,Coudreau}.

\acknowledgements We thank Igor Sokolov and Alberto Amo for
fruitful discussions. This work is supported by the EC under the
ICT/FET project COMPAS, by the RFBR (Grants No. 10-02-00103), and
by the CAPES/COFECUB project QE-COMET. O. Mishina acknowledges the
financial support from the Ile-de-France programme IFRAF and A.
Sheremet from the Foundation "Dynasty".

\appendix
\section{}\label{A}

In this appendix we present the details of the susceptibility
calculations for the six-level system presented in Fig.
\ref{fig1}. In section \ref{sec2} only a part of the results
related to the multi-$\Lambda$ system is discussed. Here we will
follow the derivation and then discuss the full solution in more
details.

First we introduce a set of slowly varying operators for the
off-diagonal elements of the atomic density matrix:
\begin{eqnarray}\label{A1}
    \sigma_{e_{F'} g}&=&\rho_{e_{F'}g}e^{i\omega_pt}
    \nonumber
    \\
    \sigma_{sg}&=&\rho_{sg}e^{i(\omega_p-\omega_c)t}\nonumber
    \nonumber
    \\
    \sigma_{ge}&=&\rho_{ge}e^{i(-\omega_c)t}
    \nonumber
    \\
    \sigma_{se}&=&\rho_{se}e^{i(\omega_p-2\omega_c)t}
    \nonumber
    \\
    \sigma_{e_{F'}e}&=&\rho_{e_{F'}e}e^{i(\omega_p-\omega_c)t}\label{A1}
\end{eqnarray}
where $F'=2,\,3,\,4$ refer to the excited states with different
angular momenta. In addition to the earlier introduced optical
coherences $\sigma_{e_{F'} g}$ on the probe field transition
(\ref{2.4}), (\ref{A1}) includes as well the coherence between the
ground states $\sigma_{sg}$, the optical coherence $\sigma_{ge}$
on the $|g\rangle$ to $|e\rangle$ transition, the coherence
$\sigma_{se}$ caused by the three-photon transition
$|s\rangle\rightarrow|e_{F'}\rangle\rightarrow|g\rangle\rightarrow|e\rangle$,
and the coherences $\sigma_{e_{F'}e}$ between exited states
$|e\rangle$ and $|e_{F'}\rangle$.

Now we present the subsequent solution of the density matrix
equations (\ref{2.6}). First we consider the zero order
equations taking into account only the control field:
\begin{eqnarray}\label{A2}
\dot{\rho}^{(0)}_{s\,s}&=&-\tau^{-1}_d\rho^{(0)}_{s\,s}+\tau^{-1}_d/2
                          +\gamma/2\sum_{F'}\rho^{(0)}_{e_{F'}e_{F'}}
                          \nonumber
                          \\
                          &&\,\,\,\,\,\,\,\,\,\,\,\,\,\,\,\,
                          -i\sum_{F'}
                          (\sigma^{(0)}_{s\,e_{F'}}\Omega^c_{e_{F'}s}
                          -\sigma^{(0)}_{e_{F'}s}\Omega^c_{s\,e_{F'}})/2
\nonumber
   \\
   \dot{\rho}^{(0)}_{g\,g}&=&-\tau^{-1}_d\rho^{(0)}_{g\,g}+\tau^{-1}_d/2
                          +\gamma/2\sum_{F'}\rho^{(0)}_{e_{F'}e_{F'}}
                          +\gamma\rho^{(0)}_{e\,e}
                          \nonumber
                          \\
                          &&\,\,\,\,\,\,\,\,\,\,\,\,\,\,\,\,
                          -i(\sigma^{(0)}_{g\,e}\Omega^c_{e\,g}
                          -\sigma^{(0)}_{e\,g}\Omega^c_{g\,e})/2
\nonumber
\\
\dot{\rho}^{(0)}_{e_{F'_1}e_{F'_2}}&=&-(i\omega_{e_{F'_1}e_{F'_2}}+\gamma)\rho^{(0)}_{e_{F'_1}e_{F'_2}}
                          \nonumber
                          \\
                          &&\,\,\,\,\,\,\,\,\,\,\,\,\,\,\,\,
                          -i(\rho^{(0)}_{e_{F'_1}s}\Omega^c_{s\,e_{F'_2}}
                          -\rho^{(0)}_{s\,e_{F'_2}}\Omega^c_{e_{F'_1}s})/2
\end{eqnarray}
\begin{eqnarray}\label{A3}
\dot{\sigma}^{(0)}_{e\,g}&=&(i\Delta_c
                                  -i\omega_{e\,e_2}
                                  -\frac{\gamma}{2})\sigma^{(0)}_{e\,g}
                                  +i(\rho^{(0)}_{g\,g}
                                  -\rho^{(0)}_{e\,e})\frac{\Omega^c_{e\,g}}{2}
\nonumber
\\
\dot{\sigma}^{(0)}_{e_{F'}s}&=&(i\Delta_c-i\omega_{e_{F'}\!e_2}
                                 -\frac{\gamma}{2})\sigma^{(0)}_{e_{F'}\!s}
                                 \nonumber
                                 \\
                          &&\,\,\,\,\,\,\,\,\,\,\,\,\,\,\,\,
                          +i\rho^{(0)}_{ss}\frac{\Omega^c_{e_{F'}\!s}}{2}
                            -i\sum_{F'_1}\rho^{(0)}_{e_{F'}\!e_{F'_1}}\frac{\Omega^c_{e_{F'_1}\!s}}{2}
\end{eqnarray}
Here notations are the same as in Sec. \ref{sec2}: the Rabi
frequencies are $\Omega^c_{e_{F'}s}=2\,d_{e_{F'}
s}\varepsilon_c/\hbar$,
$\Omega^c_{e\,g}=2\,d_{e\,g}\varepsilon_c/\hbar$,
$\Omega^p_{e_{F'}g}=2\,d_{e_{F'} g}\varepsilon_p/\hbar$ and the
transition energies are  $\omega_{ij}=(E_i-E_j)/\hbar$, where
$E_i$ is the energy of the unperturbed atomic state $|i\rangle$.
Indexes $F'$ and $F'_1$ run through the values $2$, $3$ and $4$.

Decay mechanisms included in our equations model an excited state
radiative decay with the rate $\gamma$ and a ground state decay
due to the finite time of flight $\tau_d$ of the atoms through the
light beam. We assume that the loss rate of atoms is much smaller
that the radiative decay rate $\tau^{-1}_d\ll\gamma$. For our
calculations we consider a vapor of cesium atoms. The radiative
decay is then $\gamma^{-1}=30$ ns, and we assume $\tau_d =
300\,\mu$s, corresponding to the time of flight at room temperature through a light
beam of few centimeters in diameter. We assume
that at the same rate as atoms are leaving the beam, new
unpolarized atoms will be entering the interaction region. This is
included by the source term $+\tau^{-1}_d/2$ in equations
(\ref{A2}).

Throughout the article we use the conditions that all Rabi
frequencies are much smaller than the hyperfine splittings
$\Omega^c_{e_{F'}s}\ll\omega_{e_{F'}e_2}$. In this case we neglect
the coherences $\rho^{(0)}_{e_{F'_1}\,e_{F'_2}}$ between different
excited states $F'_1\neq F'_2$, which brings us to the following
stationary solutions:
\begin{eqnarray}\label{A4}
  \rho^{(0)}_{ss}&=&\frac{1}{2}\left(1+\tau_d\,\sum_{F'}\frac{|\Omega^c_{e_{F'}\,s}|^2\gamma/2}
  {4(\Delta_c-\omega_{e\,e_2})^2+\gamma^2+|\Omega^c_{e_{F'}\,s}|^2}\right)^{-1}
  \nonumber
  \\
  \rho^{(0)}_{gg}&=&\frac{1}{2}+\tau_d\,\rho^{(0)}_{ss}\sum_{F'}\frac{|\Omega^c_{e_{F'}\,s}|^2\gamma/2}
  {4(\Delta_c-\omega_{e\,e_2})^2+\gamma^2+|\Omega^c_{e_{F'}\,s}|^2}
  \nonumber
  \\
  \sigma^{(0)}_{g\,e}&=&\frac{2(\Delta_c-\,\omega_{e\,e_2}-i\,\gamma/2)\,\Omega^c_{g\,e}}
  {4(\Delta_c-\omega_{e\,e_2})^2+\gamma^2+|\Omega^c_{e\,g}|^2}\,
  \rho^{(0)}_{gg}
\end{eqnarray}
Using the solution (\ref{A4}) it can be verified that atoms from
all the velocity classes are pumped from the state $|s\rangle$ to
the state $|g\rangle$ with a high efficiency. Note that in this six-level model all the atoms are eventually
pumped to state $|g\rangle$. Experimentally the presence of the
other hyperfine ground state with $F=4$ will affect considerably
the atomic population in state $|g\rangle$ which justifies the approximation $\rho^{(0)}_{gg}\sim 1$ made in Sec.~\ref{sec2}. Atoms from this state
$|g\rangle$ may be pumped to the state with $F=4$, which will
render them transparent for the control and probe fields. This
process represents an important limitation for the maximum optical
depth that can be achieved in the configuration of
Fig.~\ref{fig1}. In Sec.~\ref{sec5} and Appendix \ref{B} this
optical pumping mechanism will be taken fully into account. Here
we do not consider it for simplicity and to clarify the influence
of the excited hyperfine structure on the EIT signal.

Now we proceed further along Eqs.~(\ref{2.6}) towards the first
order equation with respect to the weak probe field:
\begin{widetext}
\begin{eqnarray}\label{A5}
    \frac{d\sigma^{(1)}_{e_{F'} g}}{dt}&=&(i\Delta_p
                                  -i\omega_{e_{F'} e_2}
                                  -\gamma_{e_{F'} g})\sigma^{(1)}_{e_{F'} g}
                                  +i(\rho^{(0)}_{g\,g}\Omega^p_{e_{F'} g}
                                   +\sigma^{(1)}_{s\,g}\Omega^c_{e_{F'} s}
                                   -\sigma^{(1)}_{e_{F'} e}\Omega^c_{e\,g})/2
   \nonumber
   \\
   \frac{d\sigma^{(1)}_{s\,g}}{dt}&=&(i\Delta_{p}
                                -i\Delta_{c}
                                -\gamma_{s\,g})\sigma^{(1)}_{s\,g}
                             -i\,\sigma^{(1)}_{s\,e}\Omega^c_{e\,g}/2
                             +i\,\sum^{4}_{F'=2}\sigma^{(1)}_{e_{F'} g}\Omega^c_{s\,e_{F'}}/2
   \nonumber
   \\
   \frac{d\sigma^{(1)}_{s\,e}}{dt}&=&(i\Delta_{p}
                                -2i\Delta_{c}
                                +i\omega_{s\,g}
                                +i\omega_{e\,e_2}
                                -\gamma_{s\,e})\sigma^{(1)}_{s\,e}
                             -i\,\sigma^{(1)}_{s\,g}\Omega^c_{g\,e}/2
                             +i\sum^{4}_{F'=2}\sigma^{(1)}_{e_{F'} e}\Omega^c_{s e_{F'}}/2
   \nonumber
   \\
   \frac{d\sigma^{(1)}_{e_{F'} e}}{dt}&=&(i\Delta_{p}
                                 -i\Delta_{c}
                                 +i\omega_{e\,e_{F'}}
                                 +i\omega_{s\,g}
                                 -\gamma_{e_{F'}\,e})\sigma^{(1)}_{e_{F'} e}
                            +i(\sigma^{(0)}_{g\,e}\Omega^p_{e_{F'} g}
                           -\sigma^{(1)}_{e_{F'} g}\Omega^c_{g\,e}
                           +\sigma^{(1)}_{s\,e}\Omega^c_{e_{F'} s})/2
   \\ \nonumber 
   \\ \nonumber
\end{eqnarray}
\end{widetext}
From Eqs.~(\ref{A5}) we find the steady state solution for the
optical and ground state coherences
\begin{eqnarray}\label{A6_1}
\sigma^{(1)}_{s\,g}=&&\rho^{(0)}_{g\,g}\sum_{F'}
                  \frac{\Omega^c_{s\,e_{F'}}}{2\mathbf{\Delta}_{s\,g}}
                  \left[
                  \frac{\Omega^p_{e_{F'}\,g}}{2\mathbf{\Delta}_{e_{F'}\,g}}
                  +\frac{\varepsilon_p|\Omega^c_{e\,g}|^2T'_{F'}}{4\hbar\,\mathbf{\Delta}_{e_{F'}\,e}\mathbf{\Delta}_{e_{F'}\,g}}
                  \right]
                  \nonumber
                  \\
                  &+&\sigma^{(0)}_{g\,e}\sum_{F'}\frac{\varepsilon_p\Omega^c_{e\,g}\Omega^c_{s\,e_{F'}}}{4\hbar\,\mathbf{\Delta}_{e_{F'}\,e}\mathbf{\Delta}_{s\,g}}
                  \left(
                  H'_{F'}
                  +\frac{d_{e_{F'}\,g}}{\mathbf{\Delta}_{s\,e}}
                  \right)
\end{eqnarray}
\begin{eqnarray}\label{A6_2}
\sigma^{(1)}_{e_{F'}\,g}=&-&\rho^{(0)}_{g\,g}
                            \left[
                            \frac{\Omega^p_{e_{F'}\,g}}{2\mathbf{\Delta}_{e_{F'}\,g}}
                            +\frac{|\Omega^c_{e\,g}|^2}{4\mathbf{\Delta}_{e_{F'}\,e}\mathbf{\Delta}_{e_{F'}\,g}}
                            \frac{\varepsilon_p}{\hbar}D'_{F'}
                            \right]
                  \nonumber
                  \\
                 &-&\sigma^{(1)}_{s\,g}
                 \left[
                 \frac{\Omega^c_{e_{F'}\,s}}{2\mathbf{\Delta}_{e_{F'}\,g}}
                 +\frac{|\Omega^c_{e\,g}|^2}{4\mathbf{\Delta}_{e_{F'}\,e}\mathbf{\Delta}_{e_{F'}\,g}}
                 G'_{F'}
                 \right]
                  \nonumber
                  \\
                  &-&\sigma^{(0)}_{g\,e}\frac{\Omega^c_{e\,g}}{2\mathbf{\Delta}_{e_{F'}\,e}}
                  \frac{\varepsilon_p}{\hbar}H'_{F'}
\end{eqnarray}

The denominators $\mathbf{\Delta}_{e_{F'}\,g}$,
$\mathbf{\Delta}_{e_{F'}\,e}$, $\mathbf{\Delta}_{se}$ and
$\mathbf{\Delta}_{sg}$ are presented in expression
(\ref{2.10}).

In order to clarify the structure of Eqs.~(\ref{A6_1},\ref{A6_2}), we have
grouped its terms to highlight the different interaction schemes.
The terms proportional to $\rho^{(0)}_{gg}$ with all components written explicitly are related
$\Lambda$-type interactions via the various excited levels $F'$.
This part will be dominant under the conditions we discuss in this
paper, i.e., the control field far detuned from the $|g\rangle$ to
$|e\rangle$ transition. The other terms of Eqs.~(\ref{A6_1},\ref{A6_2}) are
given as a function of the coefficients $D'_{{F'}}$, $G'_{{F'}}$,
$T'_{{F'}}$, and $H'_{F'}$, which have the following form:
\begin{eqnarray}\label{A7}
G'_{F'}&=&
                  \frac{\Omega^c_{e_{F'}\,s}}{2\mathbf{\Delta}_{e_{F'}\,g}}
                  +\frac{\Omega^c_{e_{F'}\,s}}{2\mathbf{\Delta}_{s\,e}}
                   \left(1+ \sum_{F'_1}\frac{|\Omega^c_{s\,e_{F'_1}|^2}}{4\mathbf{\Delta}_{e_{F'_1}\,e}\mathbf{\Delta}_{e_{F'_1}\,g}}
                   \right)
\nonumber
\\
H'_{F'}&=&\frac{d_{e_{F'}\,g}}{\mathbf{\Delta}_{e_{F'}g}}
     +\frac{\Omega^c_{e_{F'}\,s}}{\mathbf{\Delta}_{e_{F'}\,g}\mathbf{\Delta}_{s\,e}}
     \sum_{F'_1}\frac{\Omega^c_{s\,e_{F'_1}}d_{e_{F'_1}\,g}}{4\mathbf{\Delta}_{e_{F'_1}\,e}}
\nonumber
\\
T'_{F'}&=&
                  \frac{d_{e_{F'}\,g}}{\mathbf{\Delta}_{e_{F'}\,g}}
                  +\frac{d_{e_{F'}\,g}}{\mathbf{\Delta}_{s\,e}}
                  +\frac{\Omega^c_{e_{F'}s}}{4\mathbf{\Delta}_{s\,e}}
                   \sum_{F'_1}\frac{\Omega^c_{s\,e_{F'_1}}d_{e_{F'_1}\,g}}{\mathbf{\Delta}_{e_{F'_1}\,e}\mathbf{\Delta}_{e_{F'_1}\,g}}
\nonumber
\\
D'_{F'}&=&
                  \frac{d_{e_{F'}\,g}}{\mathbf{\Delta}_{e_{F'}\,g}}
                  +\frac{\Omega^c_{e_{F'}\,s}}{4\mathbf{\Delta}_{s\,e}}
                    \sum_{F'_1}\frac{\Omega^c_{s\,e_{F'_1}}d_{e_{F'_1}\,g}}{\mathbf{\Delta}_{e_{F'_1}\,e}\mathbf{\Delta}_{e_{F'_1}\,g}}
\end{eqnarray}
These coefficients represent the parts of the Eqs.~(\ref{A6_1},\ref{A6_2})
related to the processes involving the $|g\rangle$ to $|e\rangle$
transition. They describe then $V$-type and $N$-type processes. In
the present article we are mainly focused on the $\Lambda$-type
interactions,  which play the dominant role under the chosen
conditions. For this reason, in the main text we do not present
explicitly the terms of Eqs.~(\ref{2.9},\ref{2.10},\ref{2.11}) related
to such $V$-type and $N$-type processes. Instead we introduce the
coefficients $N_{e_{F'}g}$ and $\mathbf{\Delta}_{N}$:
\begin{eqnarray}\label{A8}
N_{e_{F'}g}&=&
                  \frac{|\Omega^c_{e\,g}|^2}
                       {4\mathbf{\Delta}_{e_{F'}\,g}}
                       \left(
                       \frac{D'_{F'}}{\mathbf{\Delta}_{e_{F'}\,e}}
                       +\frac{\Omega^c_{e_{F'}\,s}}{4\mathbf{\Delta}_{s\,g}}
                       \sum_{F'_1}
                       \frac{\Omega^c_{s\,e_{F'_1}}T'_{F'_1}}{\mathbf{\Delta}_{e_{F'_1}\,e}\mathbf{\Delta}_{e_{F'_1}\,g}}
                       \right)
\nonumber
\\
&+&\frac{|\Omega^c_{e\,g}|^2G'_{F'}}{\mathbf{\Delta}_{e_{F'}\,e}\mathbf{\Delta}_{e_{F'}\,g}}
                  \sum_{F'_1}\frac{\Omega^c_{s\,e_{F'_1}}\left(
                  d_{e_{F'_1}\,g}
                  +\frac{|\Omega^c_{e\,g}|^2T'_{F'_1}}{4\mathbf{\Delta}_{e_{F'_1}\,e}}
                  \right)}{8\mathbf{\Delta}_{s\,g}\mathbf{\Delta}_{e_{F'_1}\,g}}
\nonumber
\\
\mathbf{\Delta}_{N}&=&\sum_{F'}\frac{|\Omega^c_{e\,g}|^2\Omega^c_{s\,e_{F'}}}{2\mathbf{\Delta}_{e_{F'}\,e}\mathbf{\Delta}_{e_{F'}\,g}}
            \left(
            G'_{F'}+\frac{\Omega^c_{e_{F'}s}}{2\mathbf{\Delta}_{s\,e}}
            \right).
\end{eqnarray}

\section{}\label{B}

In this appendix we present the complete set of equations
describing the optical pumping process discussed in sec.
\ref{sec4}.

We begin with the case where a $\sigma^+$ polarized repumping
field and a $\sigma^+$ control field resonant to the atomic transitions $|F=4\rangle$ to $|F'=4\rangle$ and $|F=3\rangle$ to $|F'=2\rangle$ respectively as described in sec.
\ref{subsec5.1}. Based on the interaction Hamiltonian \ref{5.1} we
derive the following system of equations:
\begin{widetext}
\begin{eqnarray}{\label{B1}}
    \frac{d\rho_{F_an\,\,F_bn}(\Delta_D)}{dt}&=&
                                    (i\omega_{F_bF_a}-\frac{1}{\tau_d})\rho_{F_an\,\,F_bn}
                                    +\frac{\delta_{F_aF_b}}{\tau_d}\frac{f^{0}(\Delta_D)}{(2S+1)(2I+1)}
                                    +\delta_{F_aF_b}\gamma\sum_{\substack{F_a-1\leq F'\leq F_a+1\\n-1\leq k\leq n+1}}
                                    p_{|F',\,k\rangle\rightarrow |F_a,\,n\rangle}\rho_{F'_{}k\,\,F'_{}k}
                                      \nonumber
                                     \\
                                    &+&i\sum_{F'_1,\,F'_2}
                                    \left[\frac{\Omega_{F'_1n+1\,\,F_bn}\Omega_{F'_2n+1\,\,F_an}}
                                    {4\mathbf{\Delta}_{F'_1F_a}}\rho_{F'_2n+1\,\,F'_1n+1}
                                     +\frac{\Omega_{F'_1n+1\,\,F_an}\Omega_{F'_2n+1\,\,F_bn}}
                                    {4\mathbf{\Delta}^*_{F'_1F_b}}\rho_{F'_1n+1\,\,F'_2n+1}\right]
                                    \nonumber
                                    \\
                                    &-&i\sum_{F'_1,\,F_1}
                                    \left[\frac{\Omega_{F'_1n+1\,\,F_bn}\Omega_{F'_1n+1\,\,F_1n}}
                                    {4\mathbf{\Delta}_{F'_1F_a}}\rho_{F_an\,\,F_1n}
                                     +\frac{\Omega_{F'_1n+1\,\,F_an}\Omega_{F'_1n+1\,\,F_1n}}
                                    {4\mathbf{\Delta}^*_{F'_1F_b}}\rho_{F_1n\,\,F_bn}\right]
\nonumber
\\
 \frac{d\rho_{F'_an\,\,F'_bn}(\Delta_D)}{dt}&=&(i\omega_{F'_bF'_a}-\gamma)\rho_{F'_an\,\,F'_bn}
                                     -i\!\sum_{F'_1,\,F_1}
                                    \left[\frac{\Omega_{F'_bn\,\,F_1n-1}\Omega_{F'_1n\,\,F_1n-1}}
                                    {4\mathbf{\Delta}^*_{F'_aF_1}}\rho_{F'_an\,\,F'_1n}
                                    +\frac{\Omega_{F'_an\,\,F_1n-1}\Omega_{F'_1n\,\,F_1n-1}}
                                    {4\mathbf{\Delta}^*_{F'_bF_1}}\rho_{F'_1n\,\,F'_bn}\right]
                                     \nonumber
                                     \\
                                    &+&i\sum_{F_1,\,F_2}
                                    \left[\frac{\Omega_{F'_bn\,\,F_1n-1}\Omega_{F'_an\,\,F_2n-1}}
                                    {4\mathbf{\Delta}^*_{F'_aF_1}}\rho_{F_2n-1\,\,F_1n-1}
                                     +\frac{\Omega_{F'_an\,\,F_1n-1}\Omega_{F'_bn\,\,F_2n-1}}
                                    {4\mathbf{\Delta}_{F'_bF_1}}\rho_{F_1n-1\,\,F_2n-1}\right]
 \end{eqnarray}
\end{widetext}
To obtain this system we adiabatically eliminate the optical
coherences $\rho_{F'_bn+1\,\,F_an}(\Delta_D)\equiv\langle
F'_b,n+1|\rho(\Delta_D)|F_a,n\rangle$
from the full system of equations. The first equation represents
the evolution of the ground state described by the density matrix
elements $\rho_{F_a\,n\,\,F_b\,n}(\Delta_D)\equiv\langle
F_a,n|\rho(\Delta_D)|F_b,n\rangle$.
The total angular momentum of the alkali-metal in the ground state
can have two different values $I\pm1/2$, in case of the $^{133}Cs$
they will be $F_a,F_b=3,4$. The second term on the right side of
the first equation describes the source of atoms in the initial
state entering the interaction volume with a rate equal to the loss rate $1/\tau_d$. The initial population
is equally distributed among the Zeeman sublevels of the ground
states with the probability $f^{0}(\Delta_D)/(2S+1)/(2I+1)$ given
by the initial Gaussian velocity distribution
$f^{0}(\Delta_D)=(2\pi\Gamma^2_D)^{-1/2}\exp(-\Delta^2_D/2\Gamma^2_D)$
divided by the number of available ground states. By
$\delta_{i\,j}$ we define the Kronecker symbol. The last term in
the first line describes the radiative decay from different
excited states $|F',k\rangle$ with $k=n-1,n,n+1$ to the ground
state $|F_a,n\rangle$ with the rate $\gamma\cdot
p_{|F',\,k\rangle\rightarrow |F,\,n\rangle}$ given by the value of
the reduced dipole moment of the transition
$|F',\,k\rangle\rightarrow |F,\,n\rangle$ \cite{Varshalovich1988}:
\begin{equation}\label{B4}
    p_{|F',\,k\rangle\rightarrow |F,\,n\rangle} =(2J+1)(2F+1)\left[C^{F'\,k}_{F\,n\,\,1\,k-n}\right]^2%
                          \left\{\begin{array}{ccc}\frac{1}{2}&I&F\\F'&1&J\end{array}\right\}^2.%
\end{equation}
The Clebsh-Gordan coefficient $C^{F'\,k}_{F\,n\,\,1\,k-n}$
and the $6j$-symbol
$\left\{\begin{array}{ccc}\frac{1}{2}&I&F\\F'&1&J\end{array}\right\}$
give the transition properties resulting in
$\sum^{n+1}_{k=n-1}p_{|F',\,k\rangle\rightarrow |F,\,n\rangle}=1$.
The last two double sums in the first equation show the Hamiltonian dynamic in the
presence of two fields with Rabi frequencies
$\Omega_{F'_bn\,\,F_an-1}$
which contain the amplitude of the control field if $F_a=3$ or the
amplitude of the repumping filed if $F_a=4$. The denominators in these terms are equal to
$\mathbf{\Delta}_{F'_bF_a}\equiv
i\frac{\gamma}{2}-\Delta_{F'_bF_a}-\Delta_D$. Summation over the
hyperfine levels in these terms goes from $I-S-L$ to $I+S+L$ where
$S=1/2$ and $L=0$ in case of the ground states and $L=1$ in case
of the excited states.

The second equation in the system (\ref{B1}) describes the
evolution of the excited state given by the density matrix
elements $\rho_{F'_a\,n\,\,F'_b\,n}(\Delta_D)\equiv\langle
F'_a,n|\rho(\Delta_D)|F'_b,n\rangle$.
In the $D_2$-line of the alkali-metal atom, the total angular
momentum of the excited state can have four different values, in
case of the $^{133}Cs$ they will be $F'_a,F'_b=2,3,4,5$. This
equation has a structure which is similar to the first one, except
three major differences: the relaxation rate is given by the
natural decay rate $\gamma$, much larger than $1/\tau_d$; the
initial population of the excited state is zero; and there are no
source terms due to the decay from the other states.

Solving numerically equations (\ref{B1}) we obtain the velocity
distribution of atoms in state $|g\rangle\equiv|F=3,m_F=3\rangle$
modified by the optical pumping process
$f(\Delta_D)\sim\rho_{g\,g}(\Delta_D)$. It is presented in Fig.
\ref{fig6} (solid line) and used for the calculation of the
transmittance in Fig. \ref{fig7} (solid line).

As proposed in sec. \ref{subsec4.2} an additional $\sigma^+$
polarized "pump" field detuned by $\Delta_{pump}$ from the
transition $|F=3\rangle \leftrightarrow |F'=4\rangle$ can be used
to enhance the EIT contrast. The atoms with a Doppler shift
$\Delta_D=-\Delta_{pump}$ would be transferred from the state
$|g\rangle$ to the state $|F=3,m_F=3\rangle\equiv|g_3\rangle$ or
$|F=4,m_F=4\rangle\equiv|g_4\rangle$ due to a resonant optical
pumping process via the excited state
$|e\rangle\equiv|F'=4,m_{F'}=4\rangle$. To simulate this process, we
add the following terms to the right side of the first equation of
the system (\ref{B1})
\begin{eqnarray}\label{B2}
   -\gamma_{pump}(1-p_{|e\rangle\rightarrow |g\rangle})\delta_{F_a\,3}\delta_{n\,3}\,\rho_{F_a n\,\,F_a n}
   \nonumber\\
   +\gamma_{pump}p_{|e\rangle\rightarrow |g_3\rangle}\delta_{F_a\,4}\delta_{n\,3}\,\rho_{F_a n\,\,F_a n}
   \nonumber\\
   +\gamma_{pump}p_{|e\rangle\rightarrow |g_4\rangle}\delta_{F_a\,4}\delta_{n\,4}\,\rho_{F_a n\,\,F_a n}.
\end{eqnarray}
Here $\gamma_{pump}$ is an effective depopulation rate, given by expression (\ref{5.2}), and that depend on
$\Omega_{pump}$, the Rabi frequency of the pump field with respect
to the transition $|g\rangle\leftrightarrow|e\rangle$. Relative
transition rates $p_{|e\rangle\rightarrow |g\rangle}$,
$p_{|e\rangle\rightarrow |g_3\rangle}$ and
$p_{|e\rangle\rightarrow |g_4\rangle}$ take into account three
allowed transitions for the atomic decay from state $|e\rangle$ to
the states $|g\rangle$, $|g_3\rangle$ or $|g_4\rangle$
respectively. They can be found from expression (\ref{B4}) and in
case of the cesium atom $p_{e\rightarrow g}=25/60$,
$p_{e\rightarrow g_3}=7/60$ and $p_{e\rightarrow g_4}=28/60$.

\end{document}